\pdfoutput=1
\documentclass[aps,prd,reprint,superscriptaddress,longbibliography,nofootinbib]{revtex4-2}
\usepackage{graphicx}
\usepackage{amsmath}
\usepackage{amssymb}
\usepackage{amsfonts}
\usepackage{dcolumn}
\usepackage{dsfont}
\usepackage{latexsym}
\usepackage{rotating}
\usepackage{color}
\usepackage{xcolor}
\usepackage{latexsym}
\usepackage{bbm}
\usepackage{subfigure}
\usepackage{float}
\usepackage{epsfig}
\usepackage{psfrag}
\usepackage{natbib, hyperref}
\usepackage{bm}
\usepackage{amsthm}
\usepackage{siunitx}
\usepackage{eucal}
\usepackage{mathrsfs}
\usepackage{url}
\usepackage{braket}
\usepackage{ulem}
\usepackage{calligra}
\usepackage[utf8]{inputenc}
\usepackage{newunicodechar}
\usepackage{marvosym}
\usepackage[absolute]{textpos}
\usepackage[cal=cm]{mathalfa}
\usepackage{soul}
\usepackage{lipsum}

\usepackage{color}

\usepackage{hyperref}
\hypersetup{
colorlinks=true,final=true,
        linkcolor=blue,
        citecolor=blue,
        filecolor=black,
        urlcolor=blue,
}

\begin{document}
\setlength{\abovedisplayskip}{3pt}
\setlength{\belowdisplayskip}{3pt}

\title{Multiple Majorana bound states and their resilience against disorder\\ in planar Josephson junctions}

%\title{Multiple Majorana states, disorder effects and gap closing signature\\ of topological transitions in planar Josephson junctions}

%\title{Multiple Majorana bound states in planar Josephson junctions\\ with narrow superconducting leads}

\author{Pankaj Sharma}
% \email[]{pankaj@ph.iitr.ac.in}
\affiliation{Department of Physics, Indian Institute of Technology Roorkee, Roorkee 247667, India}
\author{Narayan Mohanta}
% \email[Address for correspondence:~]{narayan.mohanta@ph.iitr.ac.in}
%\email{narayan.mohanta@ph.iitr.ac.in}
\affiliation{Department of Physics, Indian Institute of Technology Roorkee, Roorkee 247667, India}

\begin{abstract}
% Planar Josephson junctions have been proposed as a platform for realizing Majorana-bound states for some years. In this work, we propose three regimes for the end states as a function of chemical potential. We show the localization of Majorana modes along with the other subgap states. We change the phase difference between the superconducting leads to tune down those subgap states close to zero energy and also show the localization of those. We also get a bigger regime at higher chemical potential values that possess a large oscillating amplitude Majorana bound state; the local density of this state shows the sharp localization towards the end of the middle metallic channel. We also show the chemical potential variation and field variation of the quasiparticle energy spectrum for the different Rashba spin-orbit coupling parameters.
Planar Josephson junctions are theoretically predicted to harbor zero-energy Majorana bound states (MBS) in a tunable two-dimensional geometry, at the two ends of the middle metallic channel. Here we show that three distinct topological superconducting regimes, governing the localization of the near-zero-energy MBS, appear in these planar Josephson junctions. The topologically-protected MBS appear near the narrow edges of the junction---not only in the middle metallic channel but also in the superconducting leads which have widths similar to the values used in recent experiments. We incorporate random fluctuation in the chemical potential to investigate the influence of non-magnetic disorder on the localization of the MBS in different topological regimes and find that the MBS are quite robust against disorder because of the two-dimensional geometry. Interestingly, moderate amount of disorder reduces the splitting between the MBS pairs, possibly by minimizing the wave function overlap of the MBS. We also discuss the changes in the topological superconducting phases  when the superconducting lead width is varied. Our results reveal a rich structure of the localization of topologically protected multiple MBS in experimentally-accessible planar Josephson junctions, and call for their experimental confirmation.
\end{abstract}
          
\maketitle

\section{Introduction}
\vspace{-0.5em}
Two-dimensional semiconductor/superconductor platforms, including the planar Josephson junctions (JJs), have been intensively investigated in recent years for the laboratory realization of zero-energy Majorana bound states (MBS)~\cite{Shabani_PRB2016, 
Hell_PRL2017, Pientka_PRX2017, 
Virtanen_PRB2018, Alidoust_PRB2018,
Stern_PRL2019, Ren_Nature2019,  Mohanta_PRApp2019,  Zhou_PRB2019,  Chen_PRL2019, Liu_PRB2019, Liu_PRB2019_2, Volpez_PRL2019, Setiawan_PRB2019, Haim_PRL2019, Melo_SciPost2019, Scharf_PRB2019,Stern_PRL2019, Eugene_2015,
Wu_PRB2020, Laeven_PRL2020, Nichele_PRL2020, Liu_PRB2020,  Glodzi_PRB2020, Volpez_PRR2020, Frolov_Nature2020,
Mohanta_CommunPhys2021, Paudel_PRB2021, Alidoust_PRB2021, Peng_PRR2021, Pakizer_PRR2021,
Grebenchuk_PRApp2022, Oshima_PRR2022, Monroe_PRApp2022,
Melo_SciPost2023, Vakili_PRB2023, Kuiri_PRB2023, Leuthi_PRB2023, Haxell_PRL2023, Banerjee_PRL2023, Banerjee_PRB2023,
Sharma_PRB2024, 
Pekerten_PRB2024, Subramanyan_PRB2024, Kuerten_2017}. The MBS are non-Abelian anyonic quasiparticles which can be used as the building blocks in topological qubits~\cite{Kitaev_2000, Ivanov_2001, Kitaev_2003, Nayak_RMP2008, Sau_PRL2010, Fu_PRL2008, Aguado_PRL2012, Oreg_PRL2010, Alicea_PRB2010, Alicea_Nature2011, Alicea_2012, Li_Nature2016, Lahtinen_SciPost2017,
 Beenaker_SciPost2020,  Zhou_NatureComm2022, Luna_SciPost2024}. These planar JJ platforms have a number of advantages over the hybrid semiconductor-superconductor nanowires, such as enhanced stability of the topological superconducting phase due to two-dimensional geometry, and good tunability with respect to chemical potential, external magnetic field and phase difference between the superconducting leads. Experimental signatures  of the MBS include the standard indicator---zero bias conductance peak, which is well-known to also originate from other mechanisms such as Andreev bound states and disorder~\cite{Suominen_PRL2017,Kuerten_2017,Ren_Nature2019}. Experimental confirmation of the topological superconducting transition via a minimum in critical supercurrent and a $0\!-\!\pi$ phase jump in ground state phase is also not free from ambiguity for realistic planar JJs, since there is no direct relationship between the critical magnetic field for topological transition and that for the extremas in the critical supercurrent or jump in the ground state phase~\cite{Sharma_PRB2024}. Despite the current experimental probing challenges, the planar JJ platforms hold a great promise for fault-tolerant topological quantum computing and a rich localization scenario of the zero-energy MBS.

\begin{figure}[t]
\begin{center}
\vspace{-0mm}
\epsfig{file=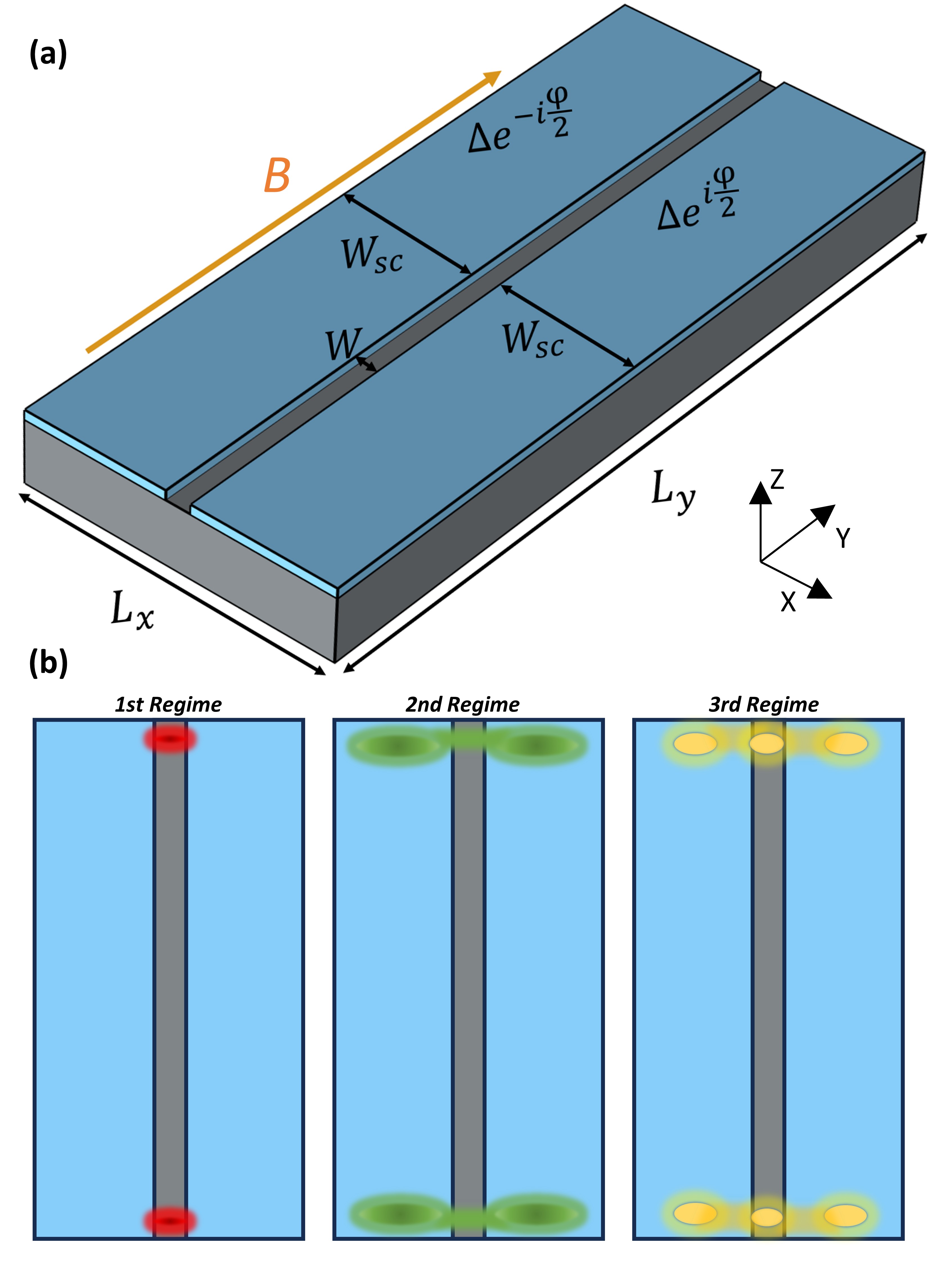,trim=0.0in 0.0in 0.0in 0.0in,clip=false, width=86mm}
\caption{(a) Schematic of a planar Josephson junction, showing two superconducting regions (in blue) deposited on a semiconductor interface (in grey) which forms a two-dimensional electron gas. A magnetic field ${\bf B}$ is applied parallel to the junction length in the entire system to tune it to the topological superconducting phase. (b) Localization of MBS in the three topological superconducting regimes.}
\label{fig1}
\vspace{-4mm}
\end{center}
\end{figure}

In this paper, we show that multiple topological superconducting transitions can occur in realistic JJs, with different localization behaviors of the zero-energy MBS. We identify three distinct topological superconducting regimes with different numbers and localization behaviors of the MBS. The first regime, with a pair of MBS localized at the two end points of the middle metallic channel, has been discussed in previous studies. The new two topological regimes, which we discuss here, host MBS also at the ends of the superconducting leads. We consider a symmetric planar JJ, with an external magnetic field applied along the length of the metallic channel, as shown in Fig.~\ref{fig1}(a). In this geometry, both the mirror symmetry with respect to the $y$-$z$ plane $M_x$ and time-reversal symmetry $\cal T$ are broken when $B\neq 0$ or $\varphi \neq 0,\pi$. But the system exhibits an effective time-reversal symmetry ${\tilde {\cal T}}=M_x \times {\cal T}$ ~\cite{Hell_PRL2017, Setiawan_PRB2019}. Hence, the system belongs to the BDI class in topological classification \cite{Schnyder_PRB2008, Tewari_PRL2012}.

Multiple topological phase transitions can occur consecutively by varying a parameter such as the chemical potential in multi-channel quantum wires, as discussed in Ref.~[\onlinecite{Eugene_2015}]. The planar JJs with thin superconducting lead width, can be compared with the case of three coupled quantum wires supporting three topological superconducting phases: the middle metallic channel and the two superconducting leads behaving like three quantum wires.
The first topological regime exhibits a pair of zero-energy MBS, localized at the ends of the metallic channel. The second topological regime hosts two pairs of low-energy MBS which are localized dominantly near the ends of the superconducting leads. The third topological regime hosts three pairs of MBS, localized at the ends of the superconducting leads and the metallic channel. The localization scenario of the MBS in the three regimes are shown schematically in Fig.~\ref{fig1}(b). The phase diagrams of the topological phases, in the plane spanned by the external magnetic field and the phase difference between the superconducting leads, reveal complex structures than a simple diamond shape found earlier~\cite{Pientka_PRX2017}. 

The comprehensive analysis presented in this paper focuses on some practical aspects of the realization of the MBS in realistic planar JJs, as outlined below. (i) Disorder in semiconductor-superconductor hybrid nanowires is known to adversely affect the MBS, and is one of the main obstacles in successful realization of the MBS in the laboratory~\cite{Ahn_PRM2021, DasSarma_PRB2021, Pan_PRB2021, Pan_PRB2022}. We show that the MBS are quite robust against non-magnetic impurities in planar JJs because of their two-dimensional nature. (ii) The nature of the localization of the MBS depends crucially on the width of the superconducting leads. The multiple MBS pairs, which we discuss here, appear in the regime of narrow superconducting leads that have been used in recent experiments~\cite{Ren_Nature2019,Fornieri_Nature2019,Dartiailh_PRL2021}.

%_____________________________________

%_____________________________________
%_____________________________________
\begin{figure*}[ht]
\begin{center}
\vspace{-0mm}
\epsfig{file=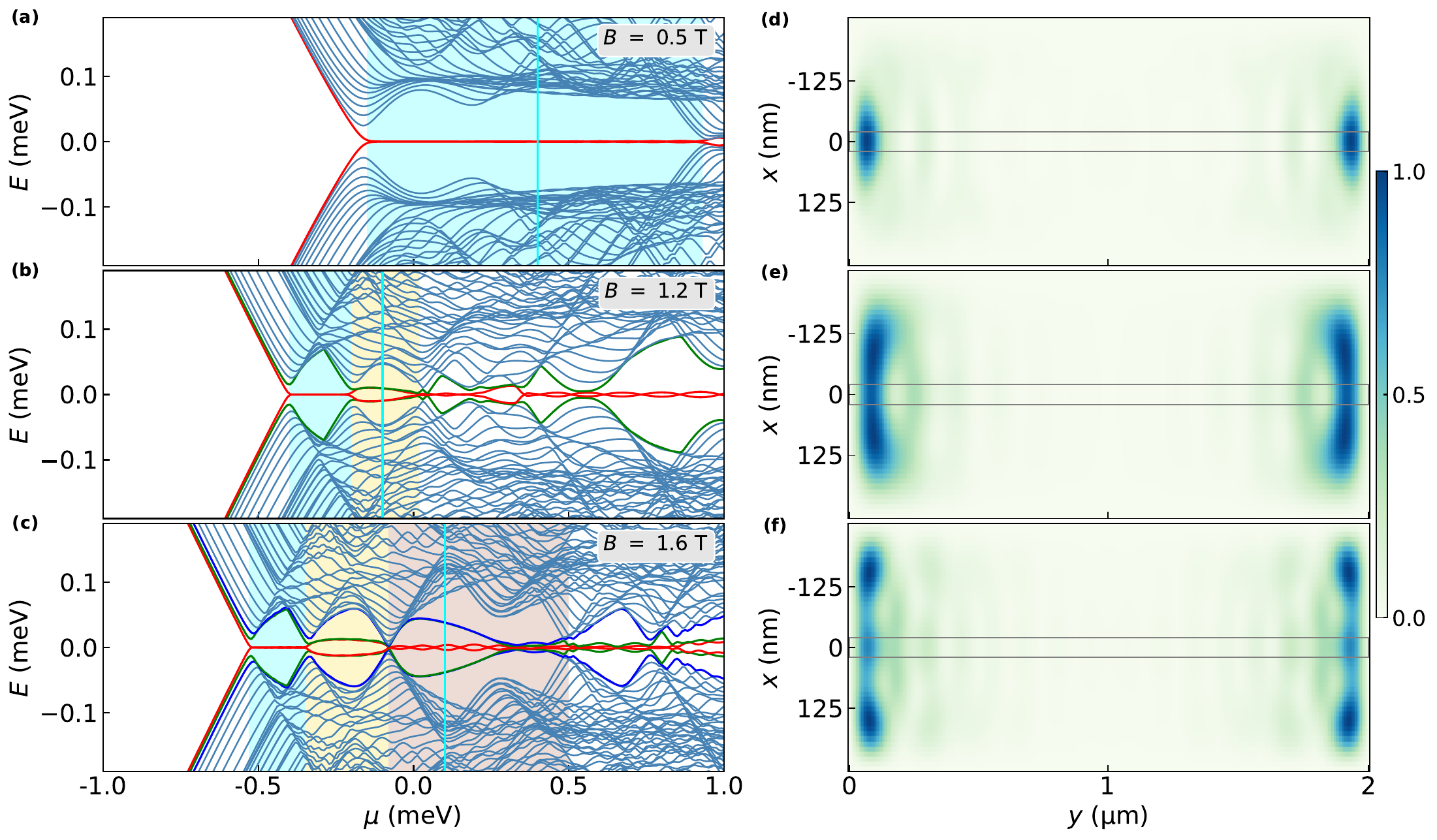,trim=0.0in 0.0in 0.0in 0.0in,clip=false, width=\textwidth}
\caption{(a)-(c) Variation of the quasiparticle energy spectrum of the planar Josephson junction with chemical potential $\mu$ at external magnetic fields $B=0.5$~T, $1.2$~T and $1.6$~T, in three identified topological regimes (shown by cyan, yellow and brown shaded regions). (d)-(f) Real-space profile of the local density of states (LDOS) of the near-zero-energy Majorana bound states at $\mu \!=\! 0.4$~meV, $-0.1$~meV, and $0.1$~meV, respectively, (shown by the cyan lines in (a)-(c)) that lie in the three topological regimes. The colorbar represent normalized LDOS corresponding to the (d) lowest positive-energy eigenstate in the first topological regime, (e) the lowest and second lowest positive-energy eigenstate in the second topological regime, and (f) the lowest three positive-energy eigenstates in the third topological regime.}
\label{fig:2}
\vspace{-4mm}
\end{center}
\end{figure*} 
%_____________________________________

The rest of this paper is organized as follows: in section II, we describe our theoretical model and methods. In section III-A, we show the multiple topological phase transitions which occur in the considered planar JJ set up, and describe phase diagrams in different planes, spanned by various control parameters. In section III-B, we analyze the stability of the MBS in different topological regimes when random non-magnetic disorder of different strengths and concentrations is present in the system. In section III-C, we discuss the variation of topological energy gap with different superconducting lead widths. In section IV, we summarize our results and discuss detection strategies for the multiple MBS.

\section{Model and method}
\vspace{0em}
The two-dimensional electron system, with proximity-induced superconductivity underneath the superconducting leads, Rashba spin-orbit coupling arising from broken inversion symmetry, and an external magnetic field applied along the metallic channel length, can be described theoretically by the below effective Hamiltonian
\begin{align} \mathcal{H}&=- t\sum_{\langle i j\rangle, \sigma} (c^\dagger_{i\sigma}c_{j\sigma}+ {\rm H.c.}) + \sum_{i, \sigma} (4t -\mu)c^\dagger_{i\sigma}c_{i\sigma}  \nonumber \\ 
&+\sum_{i } (\Delta_i c^\dagger_{i\uparrow}c^\dagger_{i\downarrow} + {\rm H.c.})   - \frac{1}{2}g \mu_{_{\rm B}}B \sum_{i, \sigma, \sigma^\prime } (\sigma_y)_{\sigma \sigma^\prime}   c^\dagger_{i\sigma} c_{i\sigma^\prime} \nonumber \\ 
&-\dfrac{\mathbf{i}\alpha}{2 a} \sum_{\langle i j\rangle, \sigma \sigma^\prime}    (\bm{\sigma} \times \mathbf{d}_{ij})^z_{\sigma \sigma^\prime}~    c^\dagger_{i\sigma}c_{j\sigma^\prime},
\label{Ham}
\end{align}
% As shown in the schematic, the Al or Pb superconducting leads deposited on the semiconductor interface create the two superconducting regions by proximity-induced superconductivity~\cite{PhysRevB.93.155402}. The below tight binding Hamiltonian describes the system under consideration
where $t\!=\!\hbar^2/2ma^2$ is the kinetic hopping energy of electrons between nearest-neighbor sites, $m$ is the effective mass of the electron, $a$ is the spacing of the considered square lattice grid, $i$ and $j$ represent lattice site indices, $\sigma $ and $\sigma^\prime$ are indices for spins ($\uparrow$, $\downarrow$), $\mu$ is the global chemical potential in the planar JJ, $\Delta_i$ is the superconducting $s$-wave pairing amplitude at site $i$, $\mu_{_{\rm B}}$ is the Bohr magneton, $g$ is the effective g-factor of electrons, $B$ is the magnetic field applied along the length of the channel ($y$ direction, as shown in Fig.~\ref{fig1}), $\bm{\sigma}$ represents the Pauli matrices, $\alpha$ is the strength of Rashba spin orbit coupling, $\mathbf{i}$ represents the imaginary number, and $\mathbf{d}_{ij}$ denotes the unit vector from site $i$ to $j$. By diagonalizing the Hamiltonian in Eq.~(1) using the unitary transformation  ${c}_{i \sigma} = \sum_{n}u^n_{i \sigma}{\gamma}_n + v^{n *}_{i \sigma} \gamma^\dagger_n$, the eigenvalues and eigenvectors are obtained, where $u^n_{i \sigma}$ ($ v^{n }_{i \sigma}$) is quasiparticle (quasihole) amplitude, and ${\gamma}_n$ (${\gamma}_n^\dagger$) is a fermionic annihilation (creation) operator of the Bogoliubov-de Gennes quasiparticles corresponding to the $n^{\rm th}$ eigenstate. The planar JJs studied in recent experiments are made of an electron gas in InAs quantum well and Al superconducting leads. Hence, we use a lattice model with the following parameters: $m\!=\!0.026m_0$, $m_0$ is the rest mass of electron, $g\!=\!10$, $\Delta \!=\!0.2$~meV, and $\alpha \!=\!30$~meV-nm~\cite{Smith_PRB1987,Mayer_APL2019,Carrad_advmat2020}. We use the lattice grid spacing $a\!=\!10$~nm. To diagonalize large matrices, we used the python packages Kwant~\cite{Groth_NJP2014} and MUMPS~\cite{Amestoy_SIAM_JMAA2001}. The phase difference $\varphi$ in the pairing amplitude $\Delta$ between the two superconducting leads is maintained at $\varphi \!=\!\pi$, throughout this paper unless mentioned otherwise. We consider planar JJs of dimensions are $L_y$ = \SI{2}{\micro \meter}, $W = 40$~nm, and $L_x$ = \SI{0.5}{\micro \meter} in results presented in subsequent section. We also thoroughly investigate the localization of the MBS in different regimes by varying the width $W_{\rm SC}$ of the superconducting leads.

%_____________________________________
\begin{figure*}[ht]
\begin{center}
\vspace{-0mm}
\epsfig{file=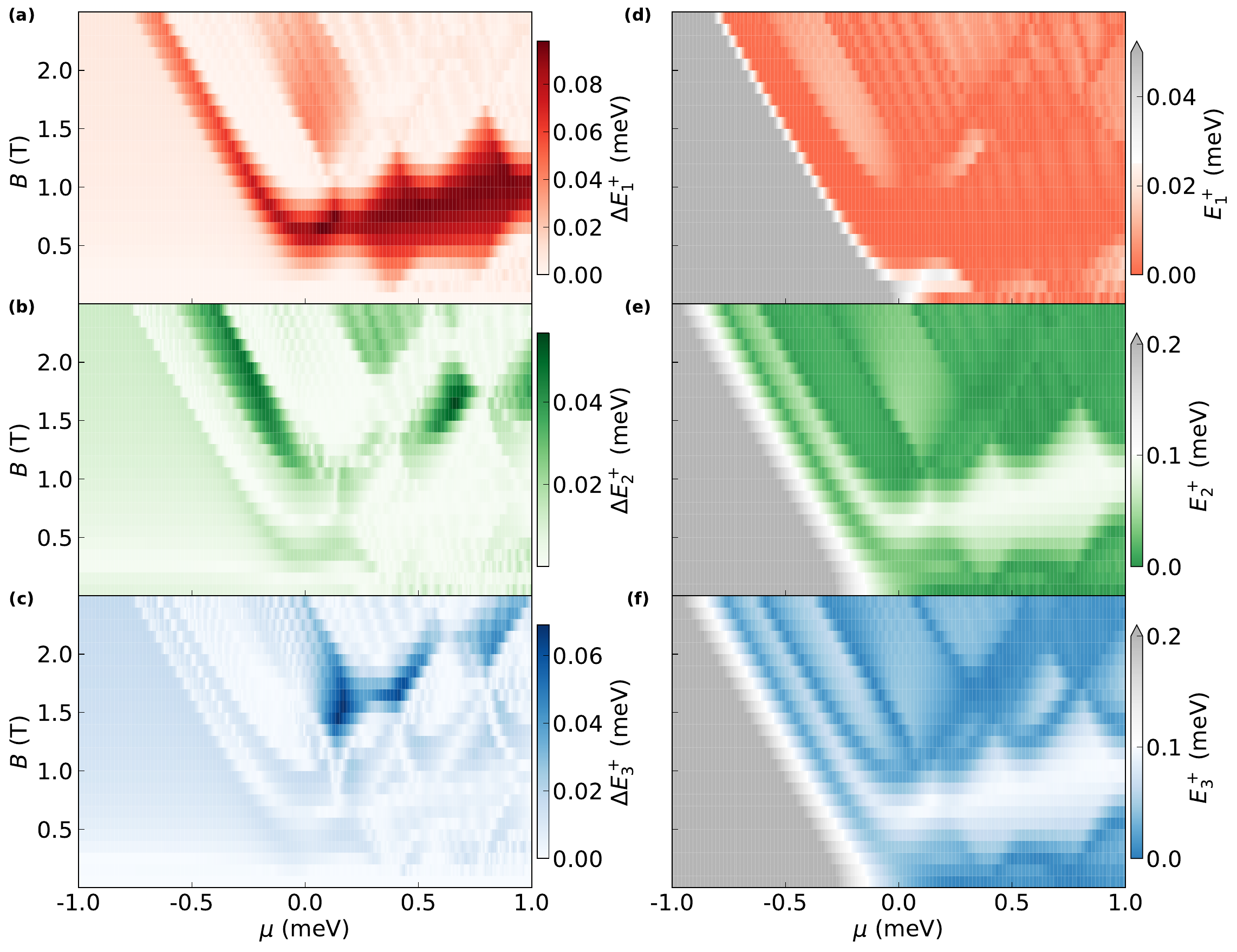,trim=0.0in 0.0in 0.0in 0.0in,clip=false, width=\textwidth}
\caption{(a)-(c) Quasiparticle energy gaps $\Delta E_{1}^+ = E_{2}^+ - E_{1}^+,~ \Delta E_{2}^+ = E_{3}^+ - E_{2}^+$ and $\Delta E_{3}^+ = E_{4}^+ - E_{3}^+$ in the plane of magnetic field ($B$) and chemical potential ($\mu$) for three topological superconducting regimes. (d)-(f) Respective quasiparticle energies $E_{1}^+$, $E_{2}^+$, and $E_{3}^+$ of the ground state, first, and second excited states.}
\label{fig:3}
\vspace{-4mm}
\end{center}
\end{figure*}
%_____________________________________

\begin{figure}[ht]
\begin{center}
\vspace{-0mm}
\epsfig{file=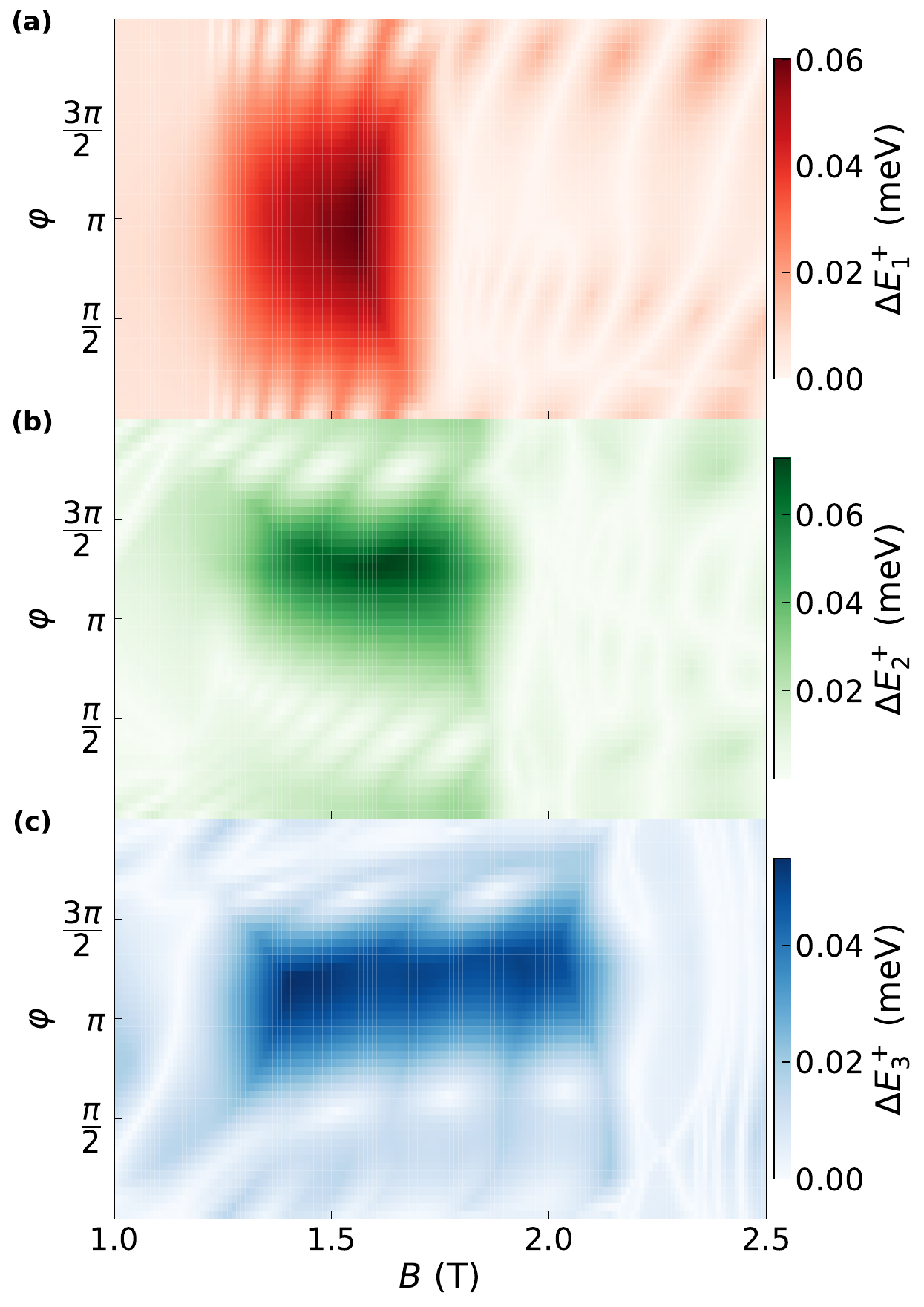,trim=0.0in 0.0in 0.0in 0.0in,clip=false, width=86mm}
\caption{Quasiparticle energy gaps $\Delta E_{1}^+ = E_{2}^+ - E_{1}^+,~ \Delta E_{2}^+ = E_{3}^+ - E_{2}^+$ and $\Delta E_{3}^+ = E_{4}^+ - E_{3}^+$ in the plane of magnetic field ($B$) and phase difference ($\varphi$) for the (a) first, (b) second, and (c) third topological superconducting regimes. These $\varphi-B$ phase diagrams were plotted at chemical potential values (a) $\mu=-0.4$~meV, (b) $\mu=-0.2$~meV, and (c) $\mu=0.1$~meV.}
\label{fig:4}
\vspace{-4mm}
\end{center}
\end{figure}
%_____________________________________

\section{Results}
\vspace{-0.5em}
\subsection{Multiple topological transitions}
\vspace{-0.5em}
We begin our efforts by identifying topological superconducting transitions in the considered planar JJ via the appearance of near-zero-energy MBS, protected by an energy gap from the bulk states. The variation of the quasiparticle energy spectrum with chemical potential $\mu$ is shown in Fig.~\ref{fig:2}(a) at a magnetic field $B\!=\!0.5~$T. The zero-energy MBS appear in the range $-0.15 \lesssim \mu \lesssim 0.93$ (in meV). The appearance of the MBS is confirmed by the local density of state (LDOS) profile, obtained via $\rho_{i} \!=\! \sum_\sigma (|u_{i \sigma}|^2 + |v_{i \sigma}|^2)$ for the lowest positive-energy eigenstate at $\mu \!=\! 0.4$~meV, shown in the Fig.~\ref{fig:2}(d), revealing localization of the MBS at the two ends of the middle metallic channel. This range of $\mu$ denotes our first topological regime. As the magnetic field is increased, the $\mu$ range within which the MBS appear at the channel ends becomes smaller, as shown in Fig.~\ref{fig:2}(b) for a higher magnetic field $B\!=\! 1.2$~T. At this field value, the first topological regime is reduced to $-0.4~{\rm meV} \lesssim \mu \lesssim -0.2$~meV; beyond this range, the pair of zero-energy MBS disappears but two pairs of topologically-protected quasiparticle states remain close to zero energy within the range  $-0.18~{\rm meV} \lesssim \mu \lesssim 0.02$~meV. In this second topological regime, the LDOS at $\mu \!=\!-0.1$~meV reveals that these near-zero-energy states are localized in the vicinity of the narrow edges of the superconducting leads, with an overlap at the metallic channel ends. With a further increase in the magnetic field, we obtain a third topological regime in which three pairs of protected states appear close to zero energy---one pair of states are similar to the usual zero-energy MBS with a typical oscillatory behavior, while the other two pairs are gapped out due to overlapping wave functions to a slightly higher energy. Such a scenario is shown in Fig.~\ref{fig:2}(c), at a magnetic field amplitude $B\!=\!1.6$~T, which gives all three topological superconducting regimes sequentially by changing $\mu$. The combined LDOS profile of the three lowest positive energy states at a chemical potential $\mu \!=\!0.1$~meV, in Fig.~\ref{fig:2}(f), shows that these states are localized at the ends of the middle metallic channel and at the ends of the superconducting leads. To confirm that the near-zero-energy protected states are MBS, we perform similar calculations on a superconducting strip, of dimension same as the superconducting leads of the considered planar JJ (see Appendix I). In the case of a quasi-two-dimensional superconducting strip, we find two similar topological superconducting regimes with multiple MBS localized at the ends of the strip. From this analysis, we conclude that the planar JJs with narrow superconducting leads can also exhibit multiple MBS localized also at the narrow edges of the leads if the magnetic field is applied in the entire JJ. In our model, the magnetic field is included uniformly throughout the entire structure. It is known, however, that in the strong-coupling regime, where the induced superconducting gap in the semiconductor is comparable to the parent superconducting gap, the effective g-factor in the proximitized region will be significantly renormalized. We found out that if we reduce the g-factor for the superconducting leads region, we can achieve multiple transitions at higher magnetic fields. We also discussed the quasi-particle eigen-energy spectrum of the planar JJ with periodic boundary condition along the channel length direction and winding number phase diagram when multiple topological transitions occur, in Appendix III and V, respectively.

%%%%%%%%%%%%%%%%%%%%%%%%%%%%%%%%%%%%%%%%%%%%%%%%%%%%%%%%%%%%%%%%%%%%%%%%%%%%%%%%%  B-mu phase diagrams

As the phase difference between the superconducting regions $\varphi = \pi$~\cite{Pientka_PRX2017} gives the maximum topological gap between the MBS and the bulk states; we set $\varphi$ at this value and calculate the $B-\mu$ phase diagrams (depicted in Fig.~\ref{fig:3}(a) - (c)) showing the quasiparticle eigenenergy differences $\Delta E_{1}^+ = E_{2}^+ - E_{1}^+,~ \Delta E_{2}^+ = E_{3}^+ - E_{2}^+$ and $\Delta E_{3}^+ = E_{4}^+ - E_{3}^+$ in red, green and blue colorplots (in the left panel) where $E_{1}^+$, $E_{2}^+$, $E_{3}^+$ and $E_{4}^+$ are the first, second, third and fourth positive energies. In Fig.~\ref{fig:3}(d) - (f), the colorplots show the quasiparticle eigenenergies $E_{1}^+$, $E_{2}^+$ and $E_{3}^+$ in the $B-\mu$ phase diagrams.
When we fix the chemical potential in the system and increase the in-plane field from zero, the first regime occurs at a critical field $B_c$. When the chemical potential is lesser, the critical field occurs at a larger value as can be seen from Fig.~\ref{fig:3}(a). The localization of the first positive energy eigenstate in this regime is towards the ends of the middle channel. It is not necessary that all the chemical potential values or the field values give the second topological regime just after the first topological regime. The second topological regime appears just after the first regime in a particular $B$-range and $\mu \lesssim 0$ meV, as can be seen from Fig.~\ref{fig:3}(b). The LDOS corresponding to the two lowest positive eigenstates for the tilted green strip in the phase diagram occurs toward the ends of the superconducting leads with some overlapping region in the middle channel. The region in the second phase diagram, starting around a chemical potential of 0.5 meV gives a mixed localization of the states in the superconducting lead as well as in the middle channel. Also, the third topological regime appears for a particular range of chemical potentials and the applied fields, as shown in Fig.~\ref{fig:3}(c). We get either the first and the second regime or the first and the third regime as we vary the in-plane magnetic field at a fixed chemical potential.

Previous theoretical calculations revealed a diamond shape topological phase on the $\varphi-B$ plane, within which the MBS exist~\cite{Pientka_PRX2017}. We calculated the $\varphi-B$ phase diagrams (presented in Fig.~\ref{fig:4}) for the multiple MBS at three values of the chemical potential: $\mu = -0.4$, $-0.2$, and $0.1$ (in meV) for the first, second, and the third regimes, respectively. The topological phase for the first regime, shown by red color in Fig.~\ref{fig:4}(a), is nearly symmetric around $\varphi=\pi$, as found earlier also~\cite{Pientka_PRX2017}. On the contrary, the topological phases for the second and the third regimes, shown by green and blue colors in Fig.~\ref{fig:4}(b)-(c), are not symmetric with respect to $\varphi=\pi$. These results reveal a rich structure of the topological superconducting phases which can occur in these planar JJ platforms in the presence of an external in-plane magnetic field. This rich structure of three consecutive topological superconducting phases is the main finding of this paper.

%The green phase diagram (presented in Fig.~\ref{fig:4}(b))  for the second regime is not symmetric around $\pi$; the protected region is shifted towards the higher values of the phase difference between the superconducting leads. The blue phase diagram (Fig.~\ref{fig:4}(c))  corresponding to the third regime looks symmetric around $\pi$ for lower magnetic field values, the topological region shifts towards the higher $\varphi$ for the higher magnetic field values.
%The $\varphi-B$ phase diagrams for the second and the third regimes reveal a richer topological region in contrast to the topological region symmetric about $\pi$ in the first regime case.

Further analysis of the planar JJ reveals that the MBS at near-zero energy in the second regime can be lowered down to zero-energy simply by changing the phase difference between the superconducting leads. (see discussion in Appendix II and related Fig.~\ref{fig:A3}).

\subsection{Influence of non-magnetic disorder on the MBS}
%%%%%%%%%%%%%% Disorder plots

\begin{figure*}[ht]
\begin{center}
\vspace{-0mm}
\epsfig{file=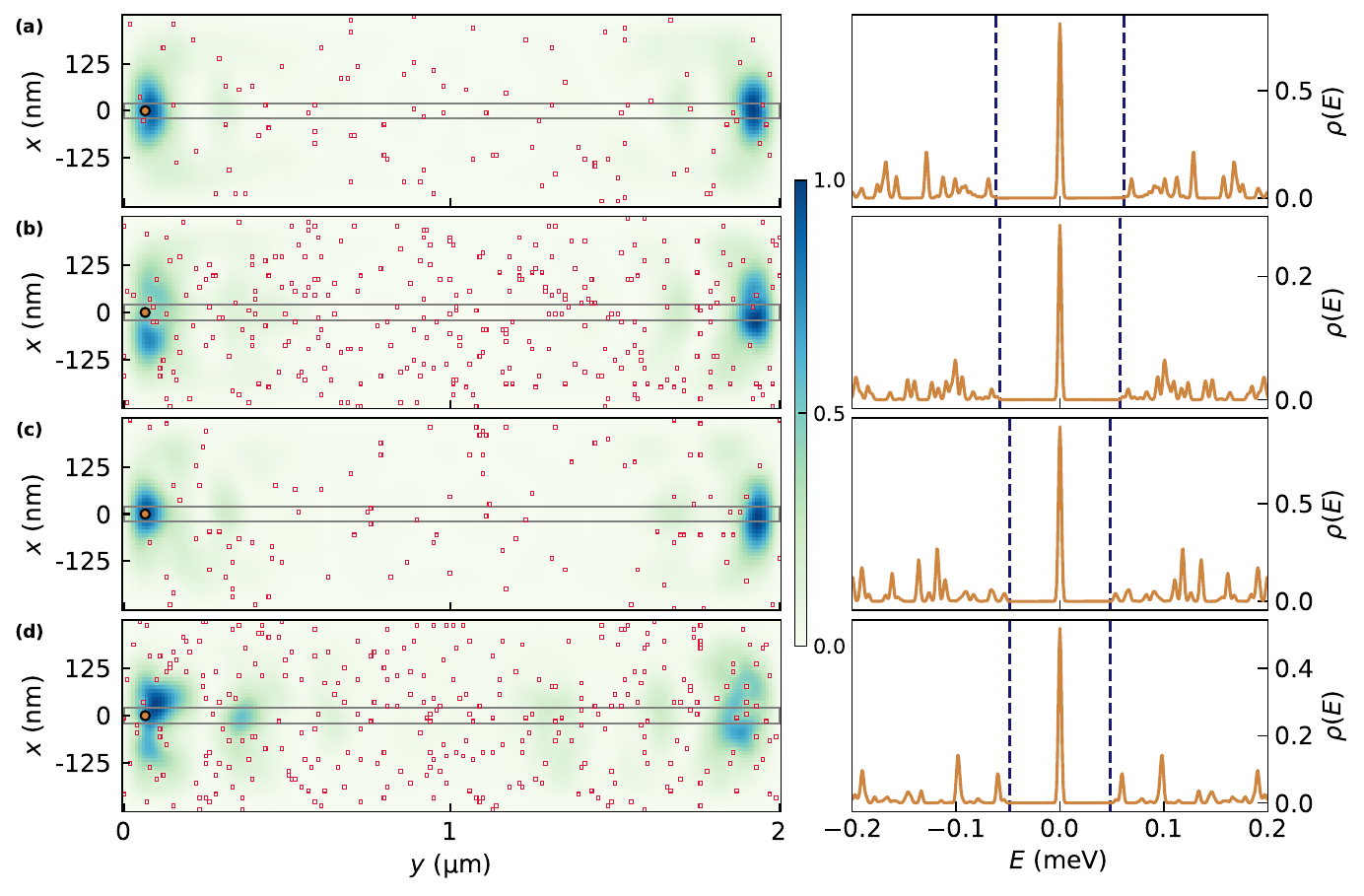,trim=0.0in 0.0in 0.0in 0.0in,clip=false, width=\textwidth}
\caption{(a),(b) Left panel: LDOS profiles for the lowest-positive-energy eigenstate at $\mu = 0.4$ meV (in the first topological superconducting regime, shown by a vertical cyan line in Fig.~\ref{fig:2}(a)) and disorder strength $W_s = 0.5t$ for (a) disorder concentration $W_c = 1\%$ and (b) $3\%$, respectively. (a),(b) Right panel:  Energy variation of the density of states at a site near the left end of the metallic channel, shown by the orange circle in (a) and (b). The plots in (c) and (d) show the observables same as in (a) and (b), with the same set of parameters, except for a larger disorder strength $W_s = t$. The red dots in the LDOS profiles represent the disordered sites. The vertical dashed lines in the right panels indicate the topological energy gap which separates the zero-energy MBS from the bulk states at higher energies.}
\label{fig:5}
\vspace{-4mm}
\end{center}
\end{figure*}
%_____________________________________

\begin{figure*}[ht]
\begin{center}
\vspace{-0mm}
\epsfig{file=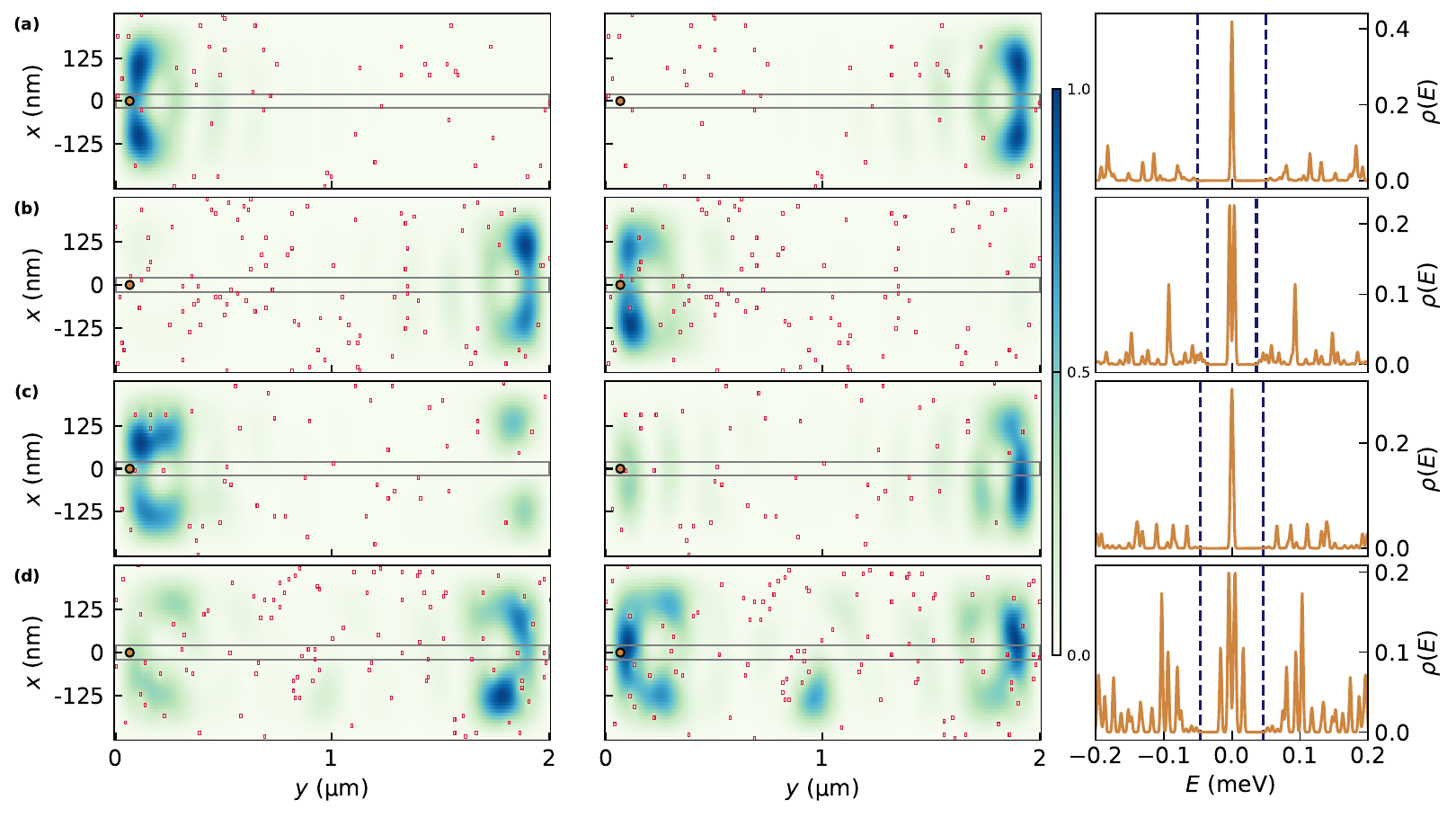,trim=0.0in 0.0in 0.0in 0.0in,clip=false, width=\textwidth}
\caption{
(a),(b): LDOS profiles for the lowest-positive-energy eigenstate (left panel) and second-lowest-positive-energy eigenstate (middle panel) at $\mu = -0.2~$meV, $B = 1.5~$T and $\varphi=1.27 \pi$ (in the second regime, shown in Fig.~\ref{fig:A2}(a)), for disorder strength $W_s = 0.5t$ and  disorder concentration (a) $W_c = 0.5\%$, (b) $W_c = 1\%$, respectively. (a),(b) Right panel:  Energy variation of the density of states at a site near the left end of the metallic channel, shown by the orange circle in (a) and (b). The plots in (c) and (d) show the observables same as in (a) and (b), with the same set of parameters, except for a larger disorder strength $W_s = t$. The red dots in the LDOS profiles represent the disordered sites. The vertical dashed lines in the right panels indicate the topological energy gap which separates the zero-energy MBS from the bulk states at higher energies.}
\label{fig:6}
\vspace{-4mm}
\end{center}
\end{figure*}
%_____________________________________
%%%%%%%%%%%%%%%%%%%%%%%%%%%%% Disorder 
%%%%%%%%%%%%%%% subsection: disorder

Fabricating ultra-clean quantum devices remains a formidable challenge in experimental physics. To explore the robustness of the multiple MBS that we uncover here against disorder, we introduce non-magnetic disorder in our theoretical model for the planar JJs, described by the Hamiltonian ${\cal H}_{\rm dis}=\sum_{i, \sigma}\mu_i c^\dagger_{i\sigma}c_{i\sigma}$, where $\mu_i$ is the local chemical potential accounting for the non-magnetic disorder, controlled by two parameters---disorder strength ($W_s$) and disorder concentration ($W_c$). This approach allows us to assess how such disorder impacts the stability of the topological superconducting phases and properties of the MBS. For this analysis, we compute the local density of states (LDOS) for the zero-energy states localized near the end of the metallic channel, using Bogoliubov-de Gennes quasiparticle amplitudes, using the relation  
\begin{align} 
\rho_{i}(E) = \sum_n (|u^\sigma_{ni }|^2 \delta(E-E_n)+ |v^\sigma_{ni}|^2 \delta(E+E_n)).
\end{align}

We study the behaviour of the MBS in the first topological regime shown in Fig.~\ref{fig:2}(a) and the two pairs of zero-energy MBS in the second topological regime shown in Fig.~\ref{fig:A3}(a).
Fig.~\ref{fig:5}(a)-(d) show the LDOS profiles at $\mu = $ 0.4 meV with the disordered sites shown by red points, in the first topological regime shown in Fig.~\ref{fig:2}(a). The disorder strength in Fig.~\ref{fig:5}(a) and (b) is $W_s\!=\!0.5t$ while in Fig.~\ref{fig:5}(c) and (d) is $W_s\!=\!t$, where $t$ is the hopping energy of the electrons. The disorder concentration in Fig.~\ref{fig:5}(a) and (c) is $W_c\!=\!1\%$ while in Fig.~\ref{fig:5}(b) and (d) is $W_c\!=\!3\%$. 
The LDOS profiles in Fig.~\ref{fig:5}(b) and (d) depict that the disorder of appropriate strength and concentration can distort the MBS or push the MBS away from the metallic channel. The site-resolved density of states plots on the right column of Fig.~\ref{fig:5}(a)-(d), obtained at a site near the left end of the middle metallic channel, shown by the orange circle in Fig.~\ref{fig:5}(a)-(d), show that, despite the disortion in the localization of the MBS, the zero-energy MBS remain robust and protected by an energy gap. The results also indicate that when the disorder strength $W_s$ is increased, by keeping the disorder concentration $W_c$ fixed, the energy gap which protects the zero-energy MBS decreases. By continuing to increase $W_s$ and $W_c$, we find that there are critical values of these numbers below which the MBS in the first topological regime remain robust against disorder. 
We compared the above results with the stripe geometry (Appendix I) and found that the planar JJ geometry is more robust against disorder than the stripe geometry with the same $W_s$ and $W_c$. We attribute this enhanced resilience to the distinct localization characteristics of the MBS in each system. In the planar JJ, the MBS are sharply localized at the ends of the narrow metallic channel, whereas in the stripe geometry, their wavefunctions are spatially extended across nearly the entire stripe width. Consequently, the compactly localized MBS in the planar JJ are less susceptible to random potential fluctuations. Our finding in narrow-lead planar JJ is consistent with recent studies on disorder effects in related systems, including a focused comparison between Majorana wires and wide-lead Josephson junctions~\cite{Paudel_2025}.

We also explore the effects of random disorder on the localization of the MBS in the second topological regime, in which the second pair of MBS can be brought to zero energy by tuning $\varphi$ as shown Fig.~\ref{fig:A3}(a). Fig.~\ref{fig:6}(a)-(d), in the left column, show the LDOS profiles at $\mu \!=\!-0.2~$meV, $B \!=\! 1.5~$T and $\varphi \! \simeq \!1.27 \pi$. The disorder strength in Fig.~\ref{fig:6}(a) and (b) is $W_s \!=\!0.5t$, while in Fig.~\ref{fig:5}(c) and (d) it is $W_s \!=\!t$. The disorder concentration in Fig.~\ref{fig:6}(a) and (c) is $W_c \!=\!0.5\%$ while in Fig.~\ref{fig:6}(b) and (d) is $W_c \!=\!1\%$.

The localization of two pairs of zero-energy MBS in the second regime at suitable $\varphi$ occurs at the ends of the middle metallic channel as well as at the ends of the superconducting leads. When non-magnetic disorder is introduced in this regime the two states get hybridized resulting in a dumbell-shaped profiles of the zero-energy MBS near both  ends of the JJ. Strong disorder causes distortion in this dumbell shape of the MBS, as shown in Fig.~\ref{fig:6}(c)-(d). When the disorder concentration $W_c$ is increased, keeping the disorder strength $W_s$ fixed, either the bulk states comes closer to zero or the zero-energy states splits away; effectively decreasing the topological gap, as evident from the right panels of Fig.~\ref{fig:6}(a) and (b) or Fig.~\ref{fig:6}(c) and (d). As in the first topological regime, the two pairs of MBS in the second topological regime remain robust below critical values of $W_s$ and $W_c$. However, we find that the MBS in the first topological regime are more resilient against disorder than those in the second regime.
Disorder can break the effective time-reversal symmetry, reducing the symmetry class from BDI to D and causing an energy splitting of the multiple MBS. We note that in Fig~\ref{fig:6}(d), two additional trivial disorder-induced bound state appear close to MBS. To distinguish a pair of MBS from the trivial low-energy bound states, the magnetic field can be varied. The MBS pair remains at or close to zero energy until a topological phase transition takes place at a critical field strength. On the other hand, the trivial low-energy bound states should move away to higher energies with increasing the field strength. By analyzing the behavior of these states with respect to a varying parameter, it is possible to distinguish the MBS from other low-energy trivial bound states~\cite{Kuerten_2017}.

We also found a higher range of chemical potentials, in which near-zero energy MBS appear with a little oscillation in the energy levels and a large topological energy gap (see Appendix IV and Fig.~\ref{fig:A6}). These MBS were found to be robust against disorder of concentration up to $\sim5\%$.

%quasiparticle eigenstate at one end (left or right) and another zero-energy quasiparticle eigenstate at another end (right or left) can be clearly seen from Fig.~\ref{fig:6}(a). Increasing the concentration of disorder while keeping disorder strength same, the topological gap decreases (compare orange panel in Fig.~\ref{fig:6}(a) and (b)). Increasing the disorder strength may distort the dumbell-shaped localization of MBS and may pushes the MBS towards the superconducting region (Fig.~\ref{fig:6}(c)). Increasing the disorder strength and concentration beyond particular values may split the two zero-energy MBS away from each other and topological gap become vanishingly small (Fig.~\ref{fig:6}(d)). We conclude from above analysis that MBS in the first regime are more robust than the MBS in the second regime, although second regime MBS are also resilient upto a considerable random disorder strength and concentration.

\begin{figure}[ht]
\begin{center}
\vspace{-0mm}
\epsfig{file=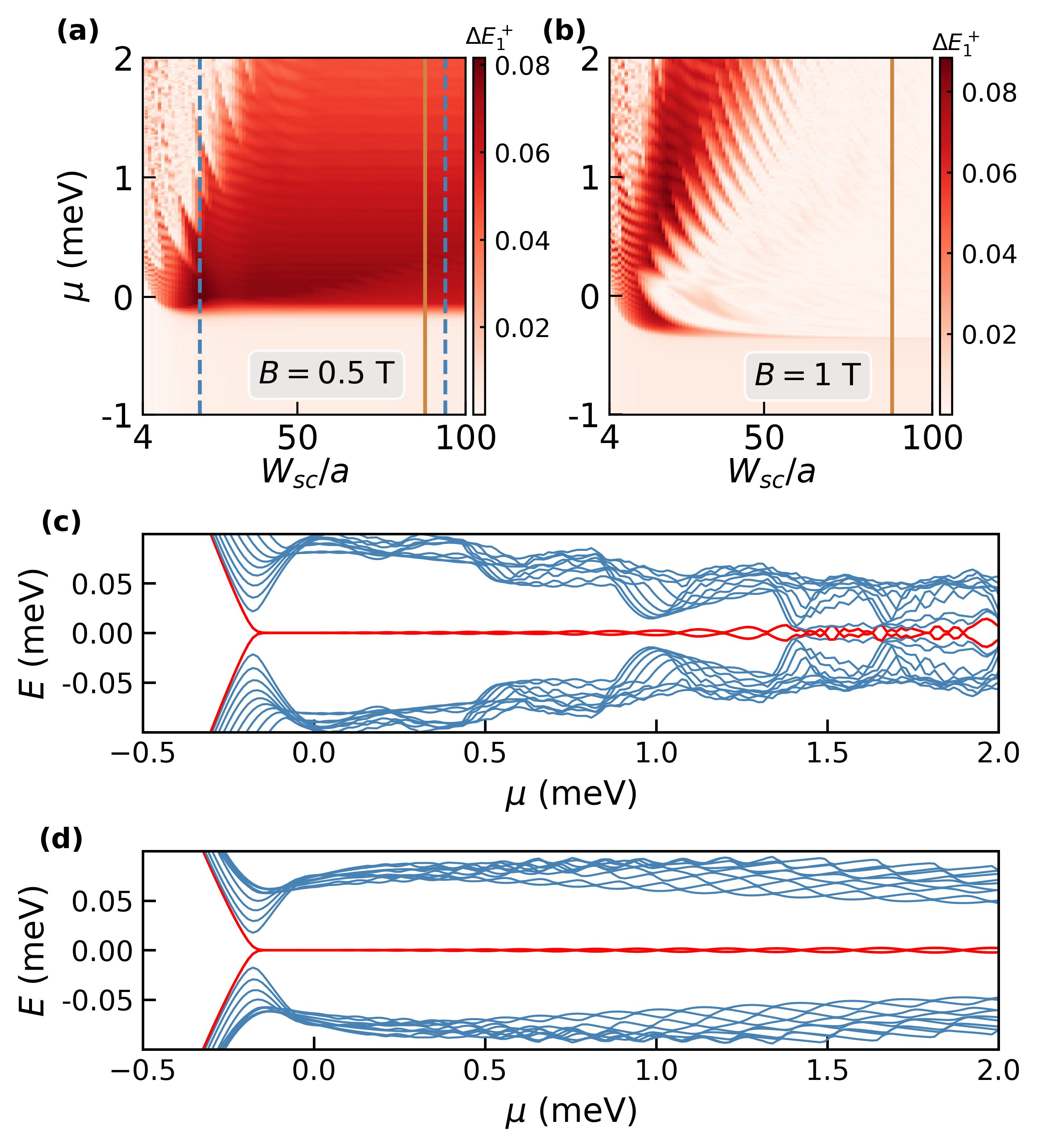,trim=0.0in 0.0in 0.0in 0.0in,clip=false, width=86 mm}
\caption{(a),(b) Energy band gap $\Delta E_{1}^+ = E_{2}^+ - E_{1}^+$, between the second-positive-energy and first-positive-energy eigenvalues, plotted in the plane of chemical potential $\mu$ and the width of the superconducting leads $W_{SC}$ for two magnetic field values (a) $B=0.5$~T, and (b) $B=1$~T in a planar JJ of dimension $L_y$=\SI{2}{\micro \meter} and $W$=\SI{0.08}{\micro \meter}. (c),(d)Variation of quasiparticle energy spectrum as a function of $\mu$ for narrow superconducting lead ($W_{SC}$=\SI{0.21}{\micro \meter}) and wide superconducting lead ($W_{SC}$=\SI{0.94}{\micro \meter}), both at $B=0.5~$T which is shown by blue dashed lines in (a). The solid vertical brown line in (a) and (b) locate the coherence length of the proximity induced superconductivity.}
\label{fig:7}
\vspace{-4mm}
\end{center}
\end{figure}
%_____________________________________
% \begin{figure}[ht]
% \begin{center}
% \vspace{-0mm}
% \epsfig{file=Fig8.jpg,trim=0.0in 0.0in 0.0in 0.0in,clip=false, width=86 mm}
% \caption{(a) - (c) shows the quasiparticle eigenenergy spectrum as a function of chemical potential for narrow superconducting lead  ($W_{SC}$ = \SI{0.21}{\micro \meter}, $W$ = \SI{0.08}{\micro \meter} and $L_{y}$ = \SI{2}{\micro \meter} at different values of in-plane magnetic field ($0.1, ~ 0.2, ~0.4$, (in units of T)). (d) - (f) show the same information but for wide superconducting lead ($W_{SC}$ = \SI{0.96}{\micro \meter}, $W$ = \SI{0.08}{\micro \meter} and $L_{y}$ = \SI{2}{\micro \meter}.}
% \label{fig:8}
% \vspace{-0mm}
% \end{center}
% \end{figure}
% %_____________________________________

\subsection{Variation of width of the superconducting leads}

The width ($W_{SC}$) of the superconducting leads in planar JJs is an important geometrical parameter which governs the nature of the topological superconducting phase. Previous theoretical studies considered both wide~\cite{Pientka_PRX2017} and narrow~\cite{Hell_PRL2017, Setiawan_PRB2019} superconducting leads of the planar JJs; however, most experimental devices fall in the regime of narrow lead width (\textit{e.g.} Refs.~\onlinecite{Ren_Nature2019,Fornieri_Nature2019,Dartiailh_PRL2021}). Planar JJs with different widths $W_{SC}$ produce different supercurrent and phase behaviors across topological superconducting transitions, which may cause difficulties in the detection of the topological phases~\cite{Sharma_PRB2024}. To show how crucially the topological superconducting phase depends on $W_{SC}$, here we investigate the energy gap $\Delta E_1^+$ between the first positive-energy state and the second positive-energy state within a range of chemical potentials, for different values of $W_{SC}$ starting from narrow-lead regime ($W_{SC} \! < \! \xi$) to wide-lead regime ($W_{SC} \! \sim \! \xi$), where $\xi \!=\!\hbar v_{F} / \pi \Delta$ $\simeq 880$~nm is the superconducting coherence length. In Figs.~\ref{fig:7}(a)-(b), we show the energy difference $\Delta E_1^+$ in the $\mu-W_{SC}$ plane, at two different values of the applied in-plane magnetic field. 
At $B = 0.5~$T, the critical chemical potential ($\mu_c$) for the topological transition is close to zero. For narrow superconducting leads ($W_{SC} < \xi$), $\mu_c$ is initially above zero but decreases gradually as the width $W_{SC}$ increases, eventually converging to a finite value beyond a certain value of $W_{SC}$. The chemical potential range over which the MBS exist broadens with increasing $W_{SC}$ at $B = 0.5~$T. This is evident from the two cases of the quasiparticle energy spectrum shown in Figs. ~\ref{fig:7}(c) and (d), which correspond to narrow and wide lead planar JJs, respectively (highlighted by the dotted lines in Fig. ~\ref{fig:7}(a)). We observe the second regime MBS around $\mu = $ 1.5 meV in Fig.~\ref{fig:7}(c). In the wide-lead planar JJ (Fig. ~\ref{fig:7}(d)), the MBS are present over a larger range of chemical potentials, accompanied by a large topological gap. 
In the $\mu - W_{SC}$ phase diagram at $B = 1 ~$T, $\mu_c$ exhibits a similar behavior as in the previous case. Here, topological gap is not present in the planar JJs with wide leads. In this case, the topological regime is predominantly observed in narrow superconducting lead planar JJs.

From the above-presented numerical results, we conclude that (i) the zero-energy MBS survive up to higher magnetic fields in planar JJs with narrower leads, (ii) planar JJs with narrow lead can host multiple MBS while we do not observe, within the considered $\mu$ range, the second bulk gap closing in planar JJs with wider leads, (iii) the zero-energy MBS appear at lower in-plane fields with a wider range of chemical potentials in the wide lead planar JJs compared to narrow lead planar JJs.

% localization of the MBS in  planar JJs. So, we explored the $\mu - W_{SC}$ phase diagrams () under different in-plane magnetic fields, which reveals a significant insights into the behavior of Majorana bound states (MBS) and the topological gap.
% In these phase diagrams, the energy difference $\Delta E_1^+$ between the ground state and the first excited state is shown, with the brown line indicating the coherence length ($\hbar v_{F} / \pi \Delta ~ \simeq 88$a)~\cite{Dartiailh_PRL2021, Fornieri_Nature2019} of the proximity-induced superconductivity. 

\section{Summary and discussion}
Our numerical investigation of the MBS in two-dimensional semiconductor/superconductor interfaces, focusing on planar JJs made of epitaxial Al-InAs systems with narrow superconducting leads, reveals a complex localization picture of topologically-protected near-zero-energy MBS. The main finding of our analysis is three successive topological superconducting regimes, consisting of one, two and three pairs of MBS, respectively, appearing when the chemical potential or the external magnetic field is varied. The MBS in the second and the third topological superconducting regimes appear at the ends of, not only the middle metallic channel, but also the two superconducting leads when the lead width is relatively smaller than their length, and it is the case in recent experimental devices. Our finding hence suggests that, in addition to the tunneling contacts at the ends of the middle metallic channel, appropriate contacts at the ends of the superconducting leads can enable the discovery of the additional low-energy MBS pairs. The successive occurrence of the three topological superconducting regimes can vouch for their topological characters, and it can serve as an identifying signature of the MBS.

We have also investigated the role of disorder in the three topological superconducting regimes discussed here. Disorder can be present in the semiconductor, in the superconductor and also at the interface between them where the localization of the MBS is assumed to be taking place dominantly. We introduce disorder in our effective two-dimensional model via random potential fluctuation, the strength and concentration of which are controlled by using two parameters. Our results suggest that the MBS are robust against disorder because of the two-dimensional geometry of the planar JJs, although when the strength and concentration are kept below some critical values. These critical values are larger in the first topological superconducting regime than the other two topological regimes. Also, in general, the MBS in planar JJs are more robust than the one-dimensional platforms such as hybrid nanowires. Remarkably, a small amount of disorder is found to reduce the energy splitting between the MBS pair by effectively minimizing the overlap of the two MBS wave functions.

To look forward, while the non-trivial fusion and braiding of the MBS, consequences of their non-Abelian properties, remain exciting challenges, continued exploration of new materials and junction designs can help to realize the topological superconducting phase without a magnetic field and to enhance the superconducting pairing gap so that the nonlocal correlation in the MBS can sustain at higher temperatures.

\section*{Aknowledgements} 
Numerical calculations were performed at the computing resources of PARAM Ganga at IIT Roorkee, provided by National Supercomputing
Mission, implemented by C-DAC, and supported by the
Ministry of Electronics and Information Technology and Department of Science and Technology, Government of India. PS was supported by Ministry of Education, Government of India via a research fellowship. NM acknowledges
a faculty initiation grant (No. IlTR/SRIC/2116/FIG)
from IITR.

%%%%%%%%%%%%%%%%%%%%%%%%%%%%%%%%%%% Appendix material
\appendix
\refstepcounter{section}
\setcounter{figure}{0}
\renewcommand{\thefigure}{A\arabic{figure}}
\renewcommand{\theHfigure}{A\arabic{figure}}
%##################################################################
\section*{Appendix I: MBS in a superconducting stripe}
%\label{appendix_I}

\begin{figure}[ht]
\begin{center}
\vspace{-0mm}
\epsfig{file=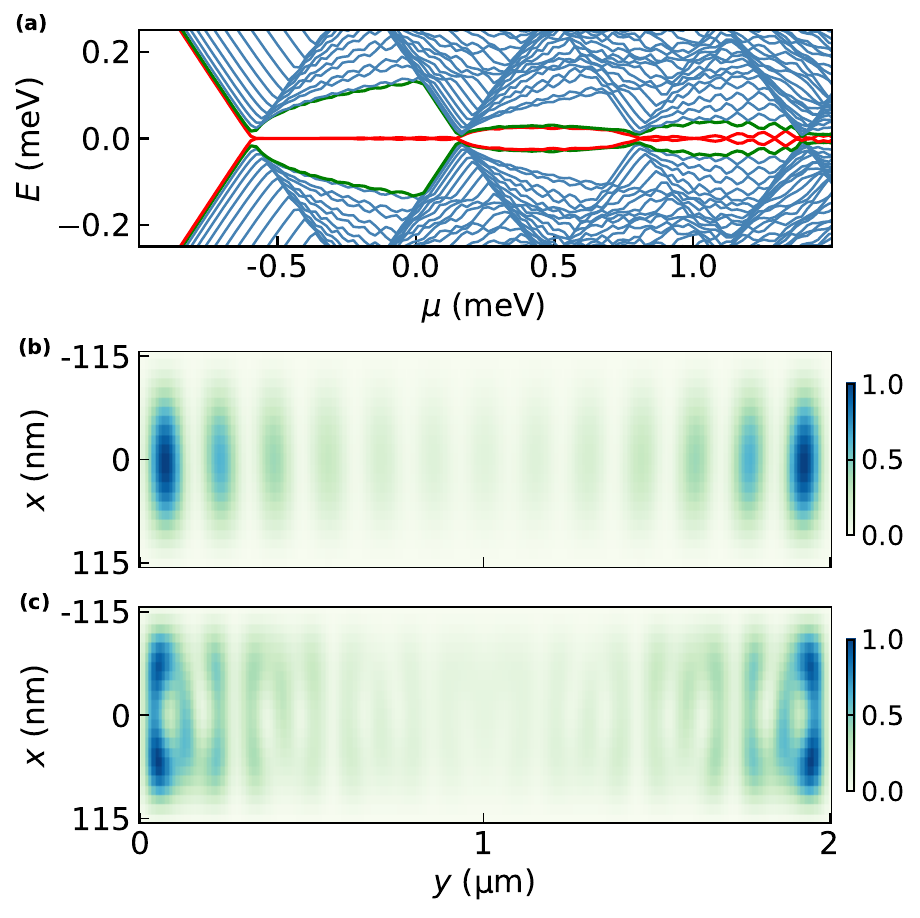,trim=0.0in 0.0in 0.0in 0.0in,clip=false, width=86mm}
\caption{(a) Chemical potential variation of quasiparticle eigenenergy spectrum of a superconducting strip with dimension $L_x = 230~$nm and $L_y$ = \SI{2}{\micro \meter}. The magnitude of the in-plane field applied along the length of the strip is $B=2.5~$T. (b) and (c) show the LDOS profiles of the superconducting strip corresponding to chemical potentials $\mu=-0.2$~meV, and $0.6$~meV, respectively.} 
\label{fig:A1}
\vspace{-4mm}
\end{center}
\end{figure}

\begin{figure*}[ht]
\begin{center}
\vspace{-0mm}
\epsfig{file=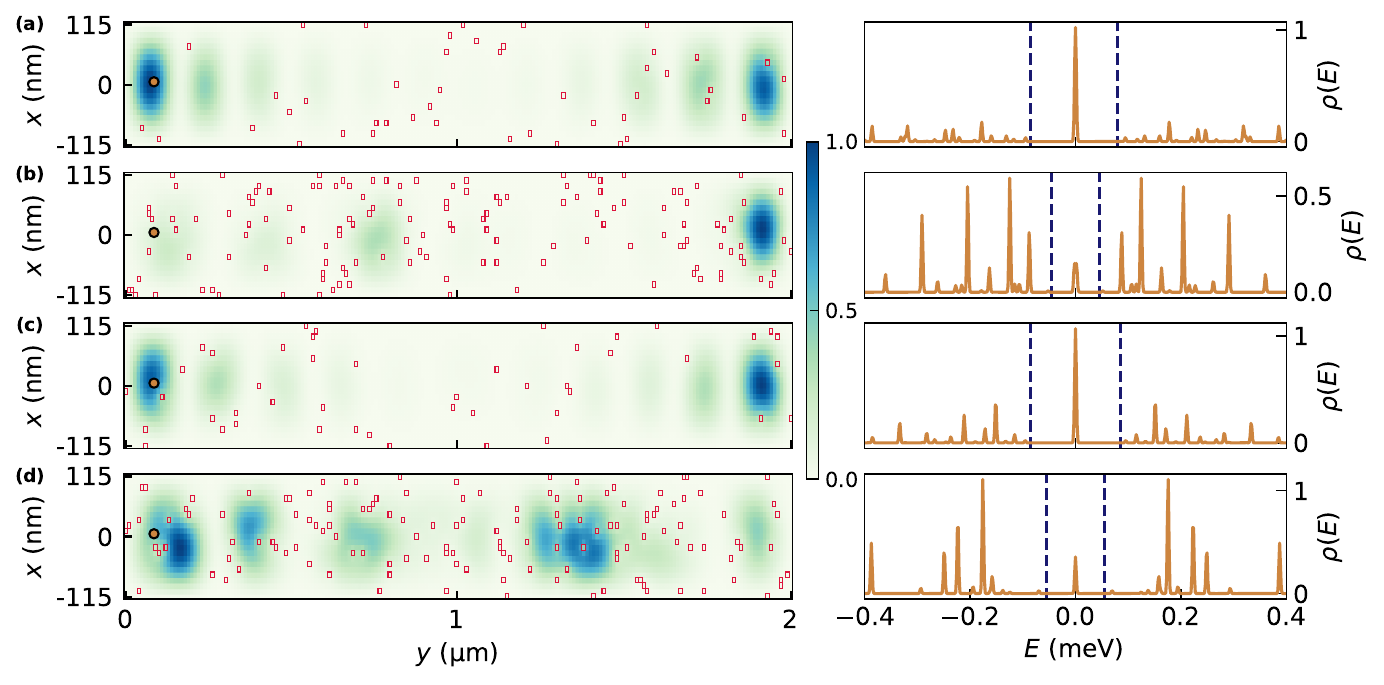,trim=0.0in 0.0in 0.0in 0.0in,clip=false, width=\textwidth}
\caption{(a),(b) Left panel: LDOS profiles for the lowest-positive-energy eigenstate at $\mu = -0.2$ meV (corresponding to Fig A1(a) ) and disorder strength $W_s = 0.5t$ for (a) disorder concentration $W_c = 1\%$ and (b) $3\%$, respectively (same as in planar Josephson junction geometry). (a),(b) Right panel:  Energy variation of the density of states at a site near the left end of the metallic channel, shown by the orange circle in (a) and (b). The plots in (c) and (d) show the observables same as in (a) and (b), with the same set of parameters, except for a larger disorder strength $W_s = t$. The red dots in the LDOS profiles represent the disordered sites. The vertical dashed lines in the right panels indicate the topological energy gap which separates the zero-energy MBS from the bulk states at higher energies.} 
\label{fig:A2}
\vspace{-4mm}
\end{center}
\end{figure*}

\begin{figure}[ht]
\begin{center}
\vspace{-0mm}
\epsfig{file=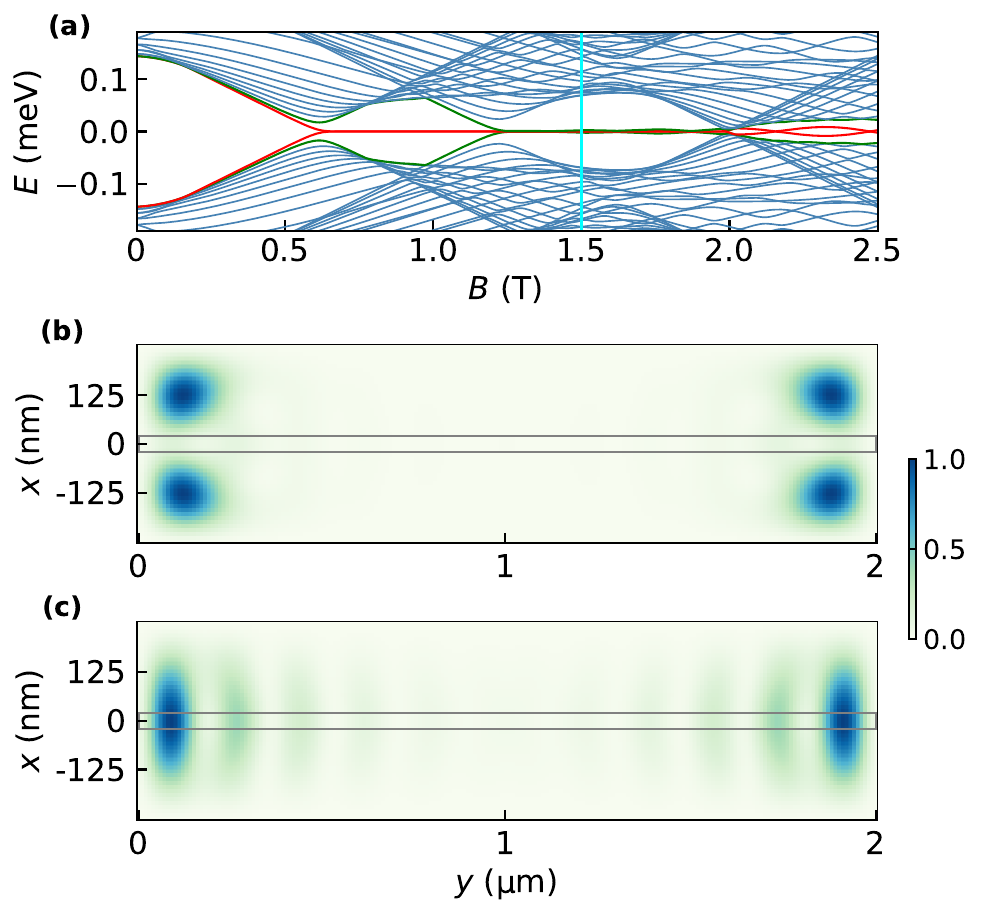,trim=0.0in 0.0in 0.0in 0.0in,clip=false, width=86mm}
\caption{(a) Magnetic field variation of the quasiparticle energy spectrum at a phase difference $\varphi \simeq 1.27\pi$ and chemical potential $\mu = -0.2~$meV. (b) and (c) show the LDOS profiles corresponding to the two lowest quasiparticle energy eigenstates at $B = 1.5~$T (indicated by the cyan line).}
\label{fig:A3}
\vspace{-4mm}
\end{center}
\end{figure}
%_____________________________________

\begin{figure}[ht]
\begin{center}
\vspace{-0mm}
\epsfig{file=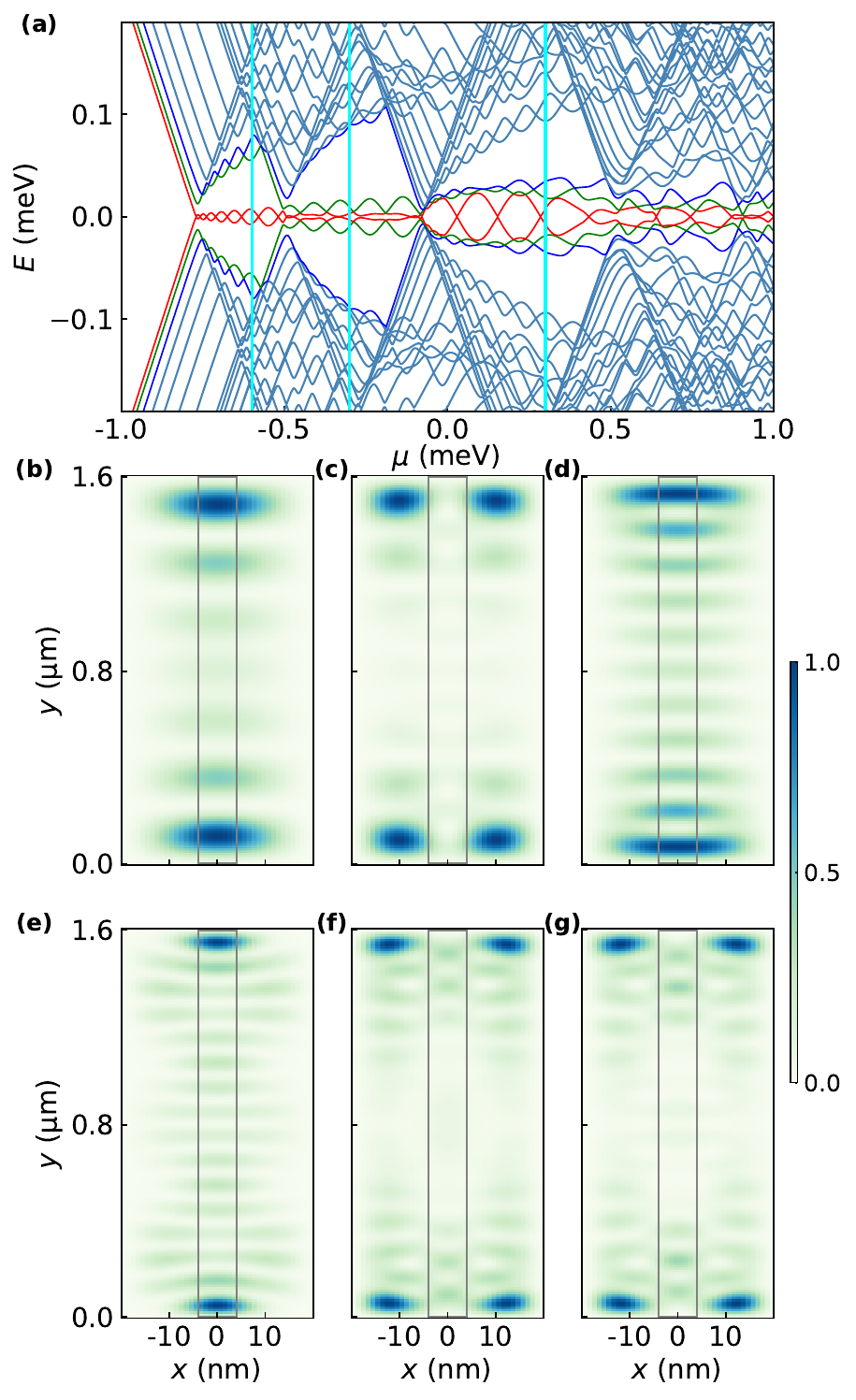, trim=0.0in 0.0in 0.0in 0.0in,clip=false, width=80mm}
\caption{(a) Chemical potential variation of quasiparticle eigenvalue spectrum of a planar JJ (of dimension $L_x = 40a$, $L_y = 160a$, $W = 8a$) at $B = 2.5$ T and $\varphi = \pi$, showing three topological superconducting regimes. (b)  The LDOS profiles correspond to the lowest-positive-energy quasiparticle at $\mu = -0.6$~meV; the two MBS are present at both ends of the channel. (c), (d) show the LDOS profiles corresponding to the lowest and second-lowest-positive energy quasiparticles, respectively, at $\mu = -0.3$~meV. (e), (f), (g) show the LDOS profiles corresponding to the lowest, second lowest and, third lowest-positive energy eigenstates, respectively, at $\mu = 0.3$ meV. }
\label{fig:A4}
\vspace{-4mm}
\end{center}
\end{figure}
%________________

\begin{figure}[ht]
\begin{center}
\vspace{-0mm}
\epsfig{file=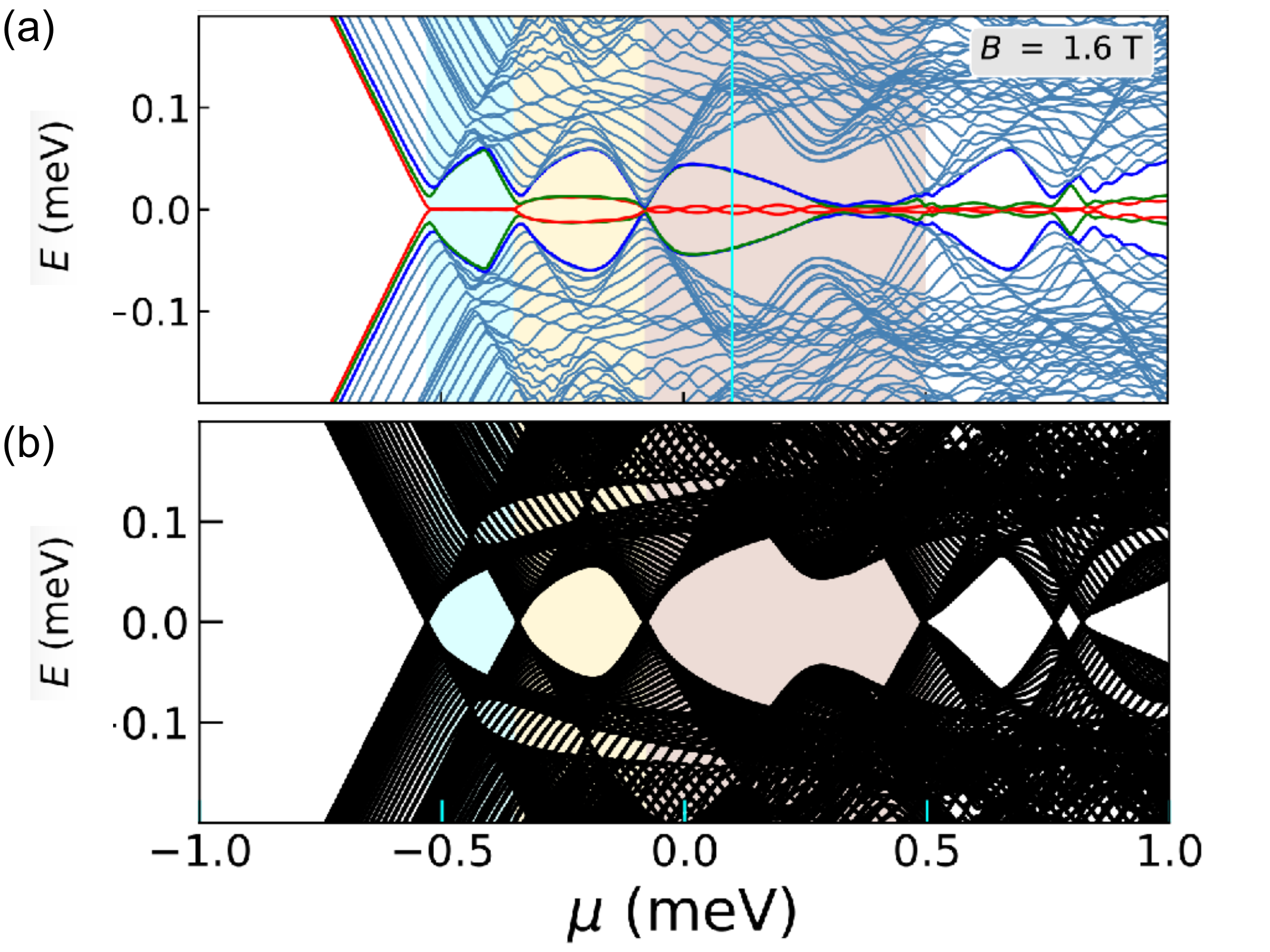,trim=0.0in 0.0in 0.0in 0.0in,clip=false, width=80mm}
\caption{ (a) and (b) shows the chemical potential variation of quasi-particle eigenenergy spectrum of a planar JJ with both finite and infinite lengths respectively having the same parameters as used in Fig 2(c).} 
\label{fig:A5}
\vspace{-4mm}
\end{center}
\end{figure}
% _____________________________________

\begin{figure*}[ht]
\begin{center}
\vspace{-0mm}
\epsfig{file=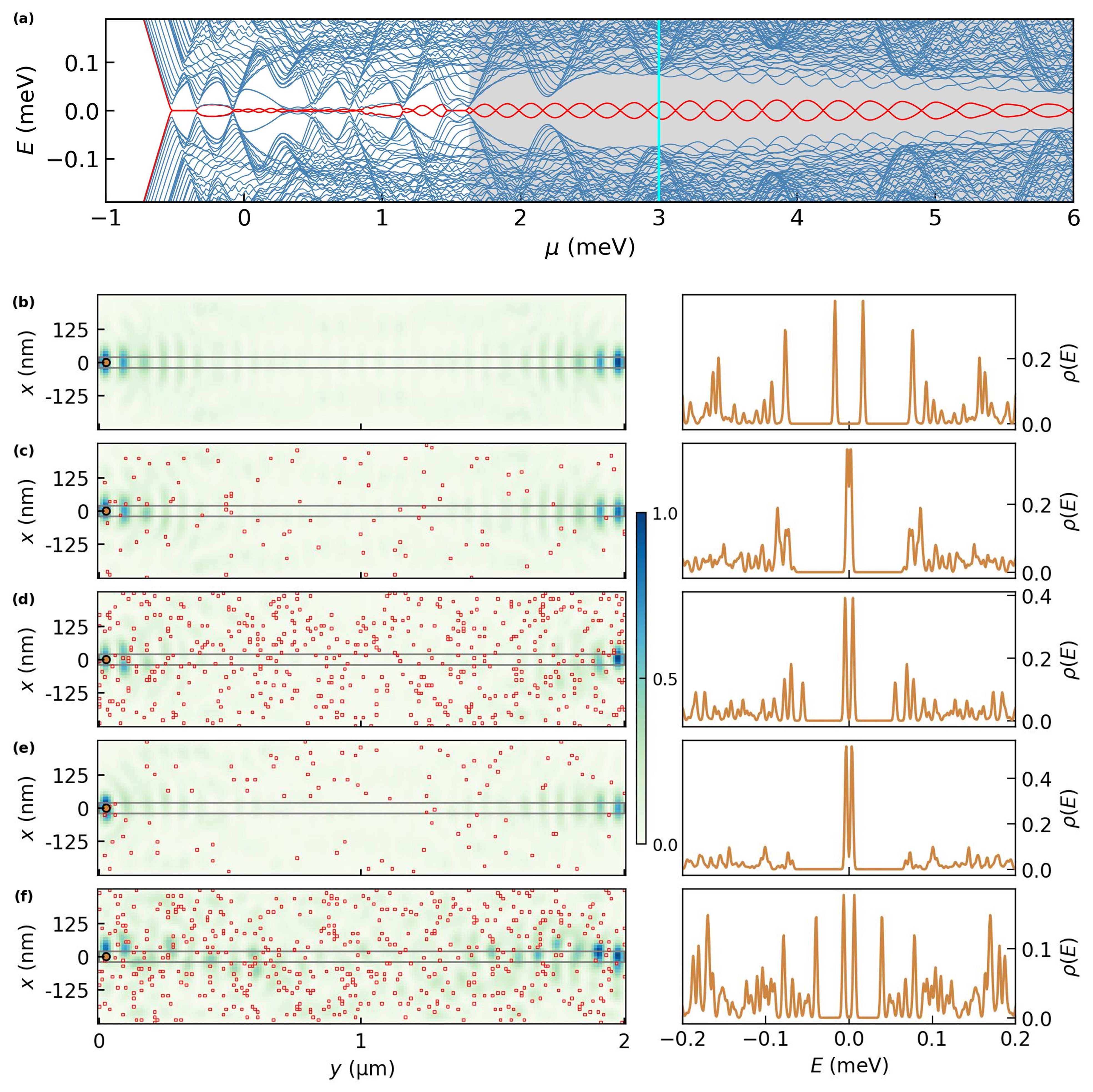, trim=0.0in 0.0in 0.0in 0.0in,clip=false, width= \textwidth}
\caption{ (a) Chemical potential variation of the quasiparticle energy spectrum at $\varphi \simeq \pi$ and $B = 1.6~$T, revealing another topological superconducting regime appearing at larger $\mu$ values (shown by the grey region). (b) Left panel: LDOS (normalized) corresponding to the lowest-positive-energy eigenstate at $\mu = 3$ meV (represented by the vertical cyan line in (a)) without any disorder, Right panel: density of state at the left end of the planar JJ revealing two MBS peaks slightly shifted from zero energy. (c) and (d) are the LDOS profiles with $W_c = 1~\%$ and $5~\% $, respectively, at a constant $W_s = 0.5t$. (e) and (f) are the LDOS profiles with $W_c = 1~\%$ and $5~\% $, respectively, at a constant $W_s = t$. The red dots in the LDOS profiles represent the disordered sites.}
\label{fig:A6}
\vspace{-4mm}
\end{center}
\end{figure*}
% _____________________________________

\begin{figure*}[ht]
\begin{center}
\vspace{-0mm}
\epsfig{file=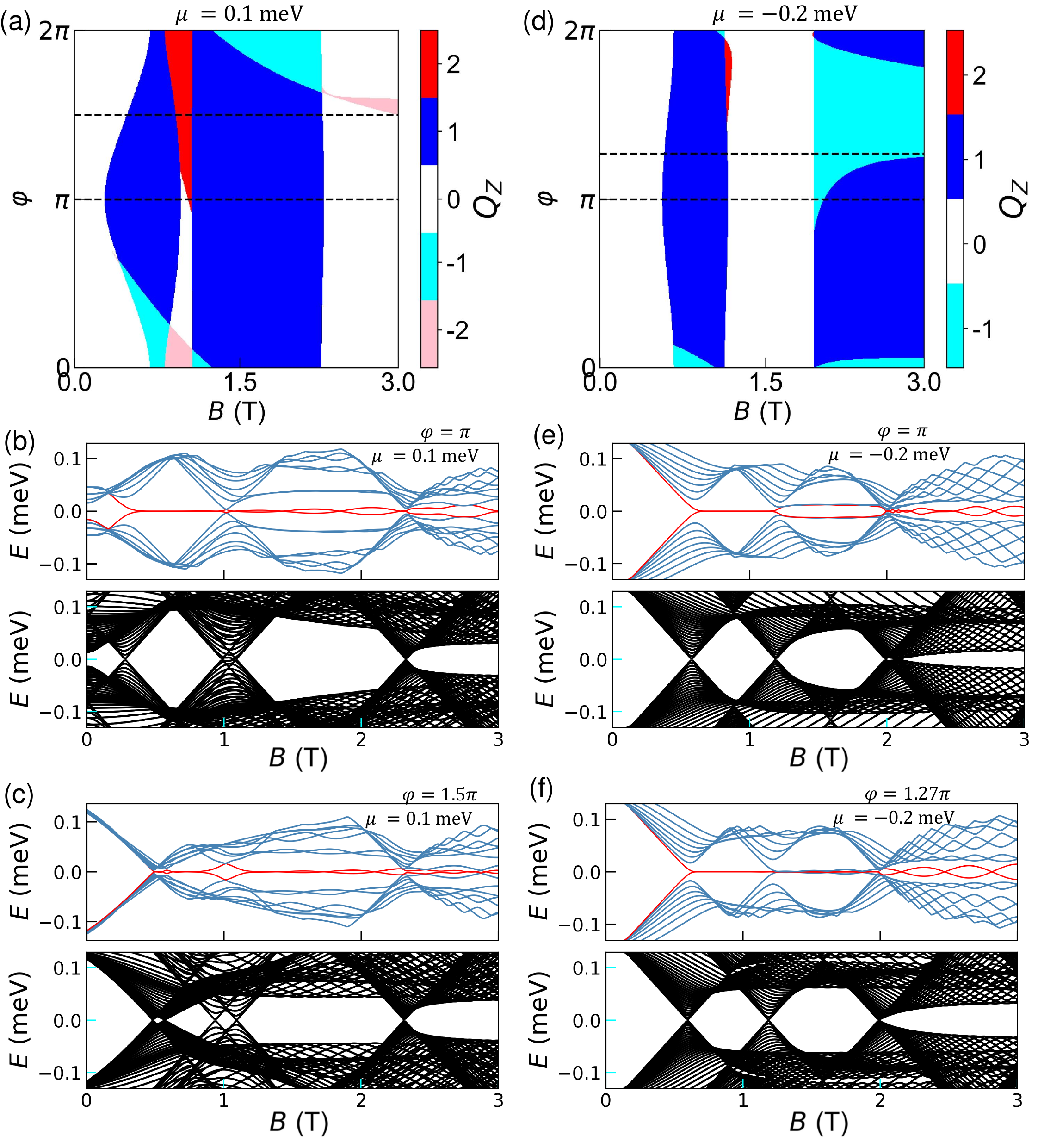, trim=0.0in 0.0in 0.0in 0.0in,clip=false, width= \textwidth}
\caption{(a) Winding number phase diagram at $\mu=0.1$ meV, using the same parameters as in the main text, except for the planar JJ length, which is considered infinite. (b) Top: Quasiparticle eigenenergy spectrum as a function of field at $\varphi = \pi$ for a finite-length planar JJ ($L_y=200a$ same as in the text). Bottom: The same spectrum for an infinite-length JJ at the same $\varphi$. (c) Same as (b), but at $\varphi = 1.5\pi$. (d) Same as (a), but at  $\mu=-0.2$ meV. (e) Same as (b), but at  $\mu=-0.2$ meV. (f) Same as (e), but at $\varphi = 1.27\pi$.}
\label{fig:A7}
\vspace{-4mm}
\end{center}
\end{figure*}

%%%%%%%%%%%%%%%%%%%%%%%%%%%%%%%%%%%%%%%
We examine a superconducting strip with dimensions identical to the superconducting lead regions in our planar JJ configuration. By applying Rashba spin-orbit coupling ($\alpha = 30$ meV-nm) and a magnetic field of approximately $2.5~$T, we get the chemical potential variation of quasiparticle eigenenergies shown in Fig.~\ref{fig:A1}(a). The resulting topological region is identified within a chemical potential range of  $-0.55 ~$meV$ \lesssim \mu \lesssim 0.1~$meV. Additionally, a second topological superconducting regime emerges, wherein the two lowest-energy states are separated from the bulk states, even though they do not reside at zero energy. Fig.~\ref{fig:A1}(b) represents the LDOS corresponding to the lowest quasiparticle eigenenergy for $\mu = -0.2~$meV.
We observe the two MBS localized towards the ends of the superconducting strip. The lower panel (Fig.~\ref{fig:A1}(c)) displays the LDOS associated with the two lowest quasiparticle eigenenergies for $\mu \!=\! 0.6$~meV in the second topological superconducting regime, where these states exhibit localization towards the strip's ends, forming a distinct dumbbell shape.

We also incorporate the random onsite potential in the stripe geometry with the same $W_{s}$ and $W_c$ as used in Fig~\ref{fig:5}, results are shown in Fig~\ref{fig:A2}. We observe that the lower concentration of the disorder does not significantly harm the localization of the MBS, depicted from (a) and (c). But as we increase the disorder concentration (shown in Fig~\ref{fig:A2}(b) and (d)), the topological gap decreases significantly and the localization of the MBS is also distorted. Whearas the localization of the MBS and topological gap remain fairly robust with the same concentration and strength of the random disorder in planar JJ geometry (Fig~\ref{fig:5}(b) and (d)). This suggests that the planar JJ geometry is more robust against the non-magnetic disorder compared to the stripe geometry. 
%The difference in these two results may arise because MBS spreads near the end sites of the strip geometry whearas the MBS in the planar Josephson junction localizes near the ends of the narrow middle mettalic channel.

\section*{Appendix II: Localization of zero-energy MBS in the second regime of planar JJ}
%\label{appendix_II}

% \lipsum[]

When analyzing the $\varphi-B$ phase diagram for the second regime, we found that at critical phase difference ($\varphi \simeq 1.27 \pi$), the energy of the two lowest-energy MBS reduces to nearly zero, beyond the first topological regime. We show this behavior by magnetic field variation of the quasiparticle energy spectrum in Fig.~\ref{fig:A3}(a).
The LDOS corresponding to the lowest and second lowest quasiparticle eigenenergies at $B = 1.5~$T with $\alpha = 30 ~$meV-nm is shown in Fig.~\ref{fig:A3}(b) and (c), respectively. Surprisingly, the lowest-energy state is localized dominantly at four locations in the superconducting regions. In contrast, the second lowest state is localized towards the end of the middle channel, similar to the localization of the usual MBS. This suggests that the localization behavior of the low-energy MBS can be controlled by changing phase difference between the superconducting regions. 

We also analyzed the chemical potential variation of the quasiparticle eigenenergy spectrum of a planar JJ having dimensions $L_x = 40a$, $L_y = 160a$, $W = 8a$ which have a comparably wider middle metallic channel shown in Fig.~\ref{fig:A4}(a). The superconducting phase difference is $\pi$ and a magnetic field of $2.5~$T is applied along the length of the junction. Three regimes are observed in ranges $-0.77 \lesssim \mu \lesssim -0.50$, $-0.50 \lesssim \mu \lesssim-0.08$ and $-0.08 \lesssim \mu \lesssim 0.50$ (in meV unit) possessing one, two and three zero or near zero energy states respectively. In the first regimes, the zero-energy MBS appear near the ends of the middle channel shown in Fig.~\ref{fig:A4}(b). The amplitude of oscillation of the MBS is significant because the length of this junction is comparably smaller than the previous geometry. Longer junction gives low amplitude MBS in the first regime.
In the second regime, we see two states oscillating at the zero-energy. One state has a larger oscillation amplitude than the other. The lowest positive energy eigenstate is localized at four positions near the ends of the superconducting regions (a surprising result also found in Fig.~\ref{fig:A3}(b)), and the second-lowest-positive-energy eigenstate is localized near the ends of the middle channel shown in Fig.~\ref{fig:A4}(c) and (d), respectively.
These calculations suggest that the dimensions of the planar Josephson junction geometry are also important parameters for realizing the two zero energy MBS after the first topological regime. 
In the third topological regime, the lowest-positive-energy eigenstate appears near the end of the channel, and the other two low-energy states appear near the ends of superconducting regions shown in Fig.~\ref{fig:A4}(e), (f) and (g) respectively, consistent with the LDOS profile in the third regime of the previous geometry (Fig.~\ref{fig:2}).

\section*{Appendix III: Planar JJ with periodic boundary conditions}

In planar JJ with infinite length, $L_y \to \infty$, there is a translational symmetry in y-direction, planar JJ model Hamiltonian in real space $\mathcal{H}(x,y)$ can be Fourier transformed using \(c_{i_x, k_y, \sigma} = \sum_{i_y'} e^{i k_y i_y' a} c_{i_x, i_y', \sigma}\) (lattice index $i \equiv (i_x, i_y)$, \(
1 \leq i_x \leq L_x \quad \text{and} \quad 1 \leq i_y \leq L_y
\) )  which gives a hybrid Hamiltonian $\mathcal{H}(x, k_y)$~\cite{Pientka_PRX2017, Hell_PRL2017, Setiawan_PRB2019}. In Fig.~\ref{fig:A5}, we show the chemical potential variation of the quasi-particle eigen-energy spectrum of the planar JJ with both finite and infinite length having same parameters. (Fig.~\ref{fig:A5}(a) is same as Fig~\ref{fig:2}(c), presented here to compare with the planar JJ with infinite length). We observe the multiple sharp gap closings around same points as in Fig~\ref{fig:2}(c) and there are no subgap-states, which means the subgap-states we observe in Fig~\ref{fig:2}(c) are due to finite length of the planar JJ.

\section*{Appendix IV: MBS at higher chemical potentials}
%\label{appendix_III}
% \lipsum[]

We also investigated the quasiparticle energy spectrum of the geometry considered in the main text at higher chemical potential at $\varphi \simeq \pi$ and $B = 1.6~T$ shown in Fig.~\ref{fig:A6}(a). We observe a broad range of chemical potentials ($1.63~meV\lesssim \mu \lesssim 6~meV$, gray region) for which MBS with larger oscillation amplitude (or splitting between the two MBS pairs) appear. This regime arises at $\mu$ values above the third topological regime. Beyond a critical $\mu$ (around $4~$meV), the oscillation amplitude of the MBS becomes nearly equal to the topological gap. As $L_y$ is increased, the oscillation amplitude in the $E-\mu$ spectrum becomes smaller. Before entering into this regime with MBS having large oscillation amplitude from the third regime, the system passes through multiple regimes having protected states, which are localized near the two ends of the planar JJ. Even though in this large-$\mu$ regime, the oscillation period of the MBS eigenvalues in $\mu$, is larger compared to first topological regime discussed above, the real-space localization profile, shown in Fig.~\ref{fig:A6}(b) left panel, shows that the wave function of the MBS are dominantly localized near the two ends of the middle metallic channel. We investigate the influence of non-magnetic disorder in this regime of MBS with larger oscillation amplitudes. In Figs.~\ref{fig:A6}(c)-(f) left panels, we show the LDOS profiles with different strengths and concentrations of disorder. The right panels of Figs.~\ref{fig:A6}(c)-(f) describe the variation of the density of states with energy $E$, at a position near the left end of the metallic channel, shown by the orange circle in Figs.~\ref{fig:A6}(c)-(f) left panel. Remarkably, with a small amount of disorder in the system, the two peaks corresponding to the MBS pair come closer towards zero energy and merges, as shown in the right panel of Fig.~\ref{fig:A6}(c). In Fig.~\ref{fig:A6}(c) and (d), the disorder strength is $W_s=0.5t$ in both LDOS configurations while the disorder concentration is  $W_c=1\%$ and  $W_c=5\%$, respectively.  In Fig.~\ref{fig:A6}(e) and (f), the disorder strength is $W_s=t$ in both LDOS configurations while the disorder concentration is  $W_c=1\%$ and  $W_c=5\%$, respectively. The topological gap decreases as the disorder concentration increases (compare right panels of Fig.~\ref{fig:A6}(c) and (d) or (e) and (f)), but even at  $W_c=5\%$, there is a considerable topological gap; shown in the right panel in Fig.~\ref{fig:A6}(d).

The above analysis of the MBS appearing in the higher chemical potential region suggests that small amount of disorder can effectively reduce the overlap of the MBS wave functions at the two metallic channel ends and help in pushing them from a finite energy towards zero energy. Also, the MBS in this regime can withstand a higher disorder concentration in comparison to the MBS in the first topological superconducting regime; the reason for this lies in the fact that the MBS in this regime are sharply localized towards the ends of the middle metallic channel.

\section*{Appendix V: BDI Phase diagrams}
We consider a planar JJ with finite length. The system's length is four times larger than its width, classifying it as a two-dimensional system instead of an effectively 1D system as considered in the literature for planar JJ with infinite length. In literature, effectively 1D or quasi-1D planar JJ, is considered to be in topological class BDI~\cite{ Pientka_PRX2017, Hell_PRL2017, Setiawan_PRB2019}, suggesting that the integer winding number ($Q_z$) is the relevant topological invariant to classify the topological superconducting phases. The Hamiltonian $\mathcal{H}(x, k_y)$ anticommutes with the chirality operator (a combination of effective time-reversal symmetry and particle-hole symmetry). In the basis where the chirality operator is diagonal (Diag($1, -1$)), the $\mathcal{H}(x, k_y)$ is block off-diagonal. The winding of the phase of the determinant of the off-diagonal part as $k_y$ changes from $0$ to $2\pi$ gives $Q_z$~\cite{Tewari_PRL2012, Setiawan_PRB2019, Pientka_PRX2017, Hell_PRL2017}.
We followed previous work ~\cite{Tewari_PRL2012, Setiawan_PRB2019} and computed the $Q_z$ invariant in the $\varphi-B$ plane for fixed chemical potentials (Figs ~\ref{fig:A7}(a) and (d)). We also calculated the quasiparticle eigenenergy spectrum for planar JJs with both finite and infinite lengths for the same parameters. We found that the two MBS pairs appear after the second bulk gap closes as the magnetic field varies but the phase diagram does not show $Q_z = 2$ in that field range (Fig ~\ref{fig:A7}(d) and (f)). Also, the $Q_z = 2$ region in the phase diagram does not show any protected state in the eigenenergy spectrum of the finite-length planar JJ (Fig ~\ref{fig:A7}(a) and (c)). We conclude in our analysis that in case of planar JJ, $Q_z$ invariant does not correctly predict the topological superconducting phases when $|Q_z| > 1$. Therefore, we use bulk gap closings, topological gap and LDOS to predict multiple MBS in the planar JJs.

\clearpage
% \newpage
%\bibliography{Ref}

%apsrev4-2.bst 2019-01-14 (MD) hand-edited version of apsrev4-1.bst
%Control: key (0)
%Control: author (8) initials jnrlst
%Control: editor formatted (1) identically to author
%Control: production of article title (0) allowed
%Control: page (0) single
%Control: year (1) truncated
%Control: production of eprint (0) enabled
%

\end{document}